\documentclass[prd,twocolumn,showpacs,preprintnumbers,nofootinbib,superscriptaddress,eqsecnum]{revtex4-1}

\usepackage{amsmath,amssymb,amsthm,color}
\usepackage{graphicx}
\usepackage{dcolumn}
\usepackage{hyperref}
\usepackage{comment}

\usepackage{color}

\newcommand {\beq}{\begin{eqnarray}}
\newcommand {\eeq}{\end{eqnarray}}
\newcommand\be{\begin{equation}}
\newcommand\ba{\begin{eqnarray}}
\newcommand\ea{\end{eqnarray}}
\newcommand\ee{\end{equation}}
\newcommand{\lb}{\left(}
\newcommand{\rb}{\right)}

\newcommand{\lsb}{\left[}
\newcommand{\rsb}{\right]}

\newcommand{\nn}{\nonumber}
\newcommand{\pd}{\partial}
\newcommand\SR{\, ^*\hspace{-0.07cm}R}
\newcommand\SF{\, ^*\hspace{-0.07cm}F}
\newcommand\aCS{\alpha_\mathrm{CS}}
\newcommand\bCS{\beta_\mathrm{CS}}
\newcommand{\mrm}{\mathrm}

\newcommand{\hhref}[1]{}

\newcommand\al{{\alpha}}
\newcommand\ep{\epsilon}

\newcommand\vt{\vartheta}

\newcommand\De{{\ensuremath{{\Delta}}}}

\newcommand\ov{\over}

\newcommand\sL{{\ensuremath{{\mathcal L}}}}

\newcommand\sO{{\ensuremath{{\mathcal O}}}}

\newcommand\sJ{{\mathcal J}}

\begin{document}
\preprint{CALT  68-2880,  IPMU14-0025, MIT-CTP 4537}

\title{Hall Viscosity and Angular Momentum in Gapless Holographic Models}

\author{Hong Liu}
\affiliation{Center for Theoretical Physics,
Massachusetts Institute of Technology,
Cambridge, MA 02139, USA}
\author{Hirosi Ooguri}
\affiliation{California Institute of Technology, Mail Code 452-48, Pasadena, CA 91125, USA}
\affiliation{Kavli Institute for the Physics and Mathematics of the Universe (WPI),\\
University of Tokyo, Kashiwa 277-8583, Japan}
\author{Bogdan Stoica}
\affiliation{California Institute of Technology, Mail Code 452-48, Pasadena, CA 91125, USA}

\date{\today}

\begin{abstract}

We use the holographic approach to compare 
the Hall viscosity $\eta_H$ and the angular momentum density
${\cal J}$ in gapless systems in $2+1$ dimensions at finite temperature. We start with a conformal fixed point and
turn on a perturbation which breaks the parity and time reversal 
symmetries via gauge and gravitational Chern-Simons couplings in the bulk.
While the ratio of $\eta_H$ and ${\cal J}$ shows some universal properties 
when the perturbation is slightly relevant, we find that 
the two quantities behave differently in general. 
In particular,  $\eta_H$ depends only on infrared physics, 
while ${\cal J}$ receives contributions from degrees of freedom at all scales.

\end{abstract}

\maketitle


\section{Introduction}

When parity and time-reversal symmetries are broken, new macroscopic
phenomena  can emerge. For example, static systems can have nonzero 
angular momenta~\cite{Volovik,stone,hoyos,sauls}, and the viscosity of energy-momentum transport can have an ``odd'' part 
(Hall viscosity) analogous to Hall conductivity~\cite{Hallviscosone}. 
In (2+1)-dimensions, these phenomena are of particular interest 
as they can occur in rotationally invariant systems.

The generation of angular momentum and of Hall viscosity are
in principle controlled by very different physics, as we discuss in detail below;
the former concerns with equilibrium thermodynamics, while the 
latter with transport. For  {\it gapped} systems at zero temperature, however, 
there exists a general argument that the two are closely related~\cite{Hallviscostwo, Hallviscostwo1} (see also \cite{Haldane,Hughes1,BGRead,CHoyosDTSon:2011ez,Hughes2,Haehl:2013kra,Son:2014}).  
In this case, the linear response of the stress tensor to an external metric perturbation 
can be described by a Berry phase, which in turn can be related to angular momentum. 
For gapless systems such an argument does not apply, and it is of interest to explore 
whether relations could exist  between the two quantities. 
Here we can take advantage of the holographic duality which provides 
a large number of strongly coupled, yet solvable, gapless systems which 
are otherwise hard to come by.

In our previous papers \cite{Liuone,Liutwo} (see also~\cite{Jensen:2011xb}) we examined 
a most general class of $(2+1)$ dimensional relativistic field theories
whose gravity description in AdS$_4$ contains axionic couplings of scalar fields to
gauge or gravitational degrees of freedom, i.e. 
\be \label{acs}
\int \vt_1  F \wedge F 
\ee
or
\be \label{bcs} 
\int  \vt_2 R \wedge R 
\ee
where $R$ is the Riemann curvature two-form, $F$ is the field strength for a bulk gauge field dual to 
a boundary $U(1)$ global current. Such terms were originally introduced in~\cite{CSgravity,Wilczek:1987,Carroll:1989vb}. In \cite{Liutwo}, we considered more general parity violating terms involving 
multiple scalar fields (see Sec.~\ref{sec:II}), but these are sufficient for illustrational purpose here. 

For convenience we will take $\vt_{1,2}$ to be odd under parity and time reversal  transformations along boundary directions
so that both~\eqref{acs} and~\eqref{bcs} are invariant under such transformations.\footnote{In top-down string theory constructions this is of course not a choice and should be determined by the fundamental theory.}  
$\vt_{1,2}$ are then dual to scalar operators $\sO_{1,2}$ in the boundary theory which are odd under
these transformations, and can be either marginal~\cite{Liuone} or relevant~\cite{Liutwo}.
The parity and time-reversal symmetries are broken when either of $\vt_{1,2}$  
is non-vanishing. In this paper we will consider models where this is achieved by turning on a scalar source or by introducing a dilatonic coupling between the scalar and gauge fields.
For both types of models, and for both~\eqref{acs} and~\eqref{bcs},  a remarkably concise and 
universal formula for the expectation value of the angular momentum density was found in terms of bulk solutions (see Sec.~\ref{sec:II} for explicit expressions).  

The Hall viscosity for holographic systems was discussed in~\cite{SaremiSon} and specific examples were presented in~\cite{Chen:2011fs,Chen:2012ti}. The result of~\cite{SaremiSon} can be readily extended  to the general class of models of~\cite{Liuone,Liutwo}~(see Sec.~\ref{sec:II}).
Thus the time is ripe for a systematic exploration of the relations between angular momentum density and 
Hall viscosity in holographic gapless systems, which will be the main goal of the current paper\footnote{See also \cite{SonWu} on the Hall viscosity and angular momentum in the holographic $p_x + i p_y$ model.  
For discussions of Hall viscosity and of other parity-violating physics in holographic as well as in field theoretic settings see \cite{Nicolis:2011ey,modulatedone,modulatedtwo,anomalyone,anomalytwo}. }.

Another motivation of the paper is that  the results of~\cite{Liuone,Liutwo} and~\cite{SaremiSon} are expressed 
in terms of abstract bulk gravity solutions, which do not always have immediate boundary interpretations.
 It should be instructive to obtain explicit expressions/values  in some simple models.

An immediate result is that, while the angular momentum density receives contributions from both the gauge and gravitational Chern-Simons terms, the Hall viscosity is only induced by the gravitational Chern-Simons term~\eqref{bcs}. That is, with $\vt_1 \neq 0$, $\vt_2 = 0$, while the angular momentum density is non-zero, the Hall viscosity vanishes. 
Moreover, even when $\vt_2\neq 0$, the Hall viscosity vanishes when the operator dual to $\vt_2$ is
marginal. This is because the holographic expression for the Hall viscosity is proportional to the
normal derivative of $\vt_2$ at the horizon. In order for it to be no-zero, some energy scale must be generated.

Thus the totalitarian principle,``everything not forbidden is compulsory,'' appears to not be at work for the Hall viscosity.  It should be emphasized that the results of~\cite{Liuone,Liutwo} and~\cite{SaremiSon}  were obtained in the classical gravity limit, which corresponds to the large $N$ and strong coupling limit of the boundary theory. The vanishing of Hall viscosity is likely a consequence of the large $N$ limit, i.e. a non-vanishing answer may emerge by taking into account loop effects on the gravity side.
In any case, it appears safe to conclude that at least for 
a certain class of holographic gapless systems, the angular momentum density and Hall viscosity 
appear not correlated at all.

The holographic expressions for the angular momentum density and  Hall viscosity 
also suggest that they are controlled by different physics.
The angular momentum density receives contributions from the full bulk spacetime, which translates 
in the boundary theory to the angular momentum density involving physics at all scales. That is, as an equilibrium thermodynamic quantity, the angular momentum behaves more like the free energy or energy, rather than entropy which depends only on IR physics. In contrast, the Hall viscosity is expressed in terms of the values of the bulk fields at the horizon, and thus depends only on IR physics.

Nevertheless it is of interest to explore whether there exist some gapless systems  or kinematic regimes 
where the angular momentum and Hall viscosity are correlated. In particular, as reviewed in Sec.~\ref{sec:II}, 
the expression for holographic angular momentum separates naturally into a sum of a contribution from the 
horizon $\sJ_{\rm horizon}$ and a contribution $\sJ_{\rm integral}$ from integrating over the bulk spacetime. 
Such a separation suggests that these contributions may have different physical origins. 
Indeed in~\cite{Liutwo} we showed that, when the operators dual to $\vt_1$ and $\vt_2$ are marginal, 
$\sJ_{\rm horizon}$  is related to anomalies ($\sJ_{\rm integral}$ vanishes in the marginal case).
Given that both $\sJ_{\rm horizon}$ and the Hall viscosity $\eta_H$ 
only involve horizon quantities, it is then natural to ask whether we could find some connection 
between them. Interestingly in various classes of models we do find the two are related in a rather simple 
way in the limit that the symmetry breaking effects are small, suggesting a possible common 
physical mechanism underlying both.  

After completing this project and while preparing this manuscript, we have received the paper \cite{Wu}, which 
disagrees with some of the results in this paper as well as those in our earlier papers \cite{Liuone,Liutwo}.
In particular, \cite{Wu} claims that the gauge Chern-Simons term does not contribute to the angular momentum density, in disagreement with  (28) and (30) of \cite{Liuone} and (2.32) and (2.50) of \cite{Liutwo},
as well as with (6.15) and (7.5) of  \cite{Jensen:2011xb}. The difference can be traced to boundary conditions at the horizon. In \cite{Liuone,Liutwo} and in this paper, we chose the time-space components of the metric to vanish at the horizon,
\begin{equation}
h_t^i=0,
\label{bcathorizon}
\end{equation}
where $i$ is in the spatial direction along the boundary. On the other hand, 
\cite{Wu} left $h_t^i$ to be arbitrary at the horizon. We chose to impose the condition (\ref{bcathorizon})
to avoid a conical sigularity
when we analytically continue $t$ to Euclidean time.

The plan of the paper is as follows. In Section II, we will summarize the holographic formulae of the angular momentum~\cite{Liuone,Liutwo} 
and of the Hall viscosity~\cite{SaremiSon}.  In section III, we will apply these formulae to discuss a class of holographic RG flows at a finite temperature
and a chemical potential, where the parity and time reversal symmetries are broken by a source for the scalar field. We will start with 
analytical results in the limit where the symmetry breaking perturbation is small and then present numerical results. 
In Section IV, we will discuss models where these symmetries are broken by the dilaton coupling to the gauge field. We will summarize our result in Section V.
In Appendix A, we will describe an analytic solution for the scalar field in the bulk near criticality. 

\section{Review of holographic angular momentum and Hall viscosity} \label{sec:II}


Here we first review the results of~\cite{Liutwo} and~\cite{SaremiSon} on angular momentum and Hall viscosity. 

We consider the most general bulk Lagrangian for the Einstein gravity with gauge and gravitational Chern-Simons terms (axionic couplings), and allow any number of Abelian gauge fields $A_{a}^P$ ($a= 0,1,2,3$; $P=1, \dots, N$) and scalar fields $\vartheta^I$ ($I=1,\dots, M$),
\be
\mathcal{L}=\frac{1}{2\kappa^2} \sqrt{-g}\lsb \mathcal{L}_0 + \mathcal{L}_{CS} \rsb,
\label{totalLag}
\ee
where $\mathcal{L}_0$ contains the Einstein-Hilbert term, the kinetic and potential terms for the scalar fields, and the Maxwell term for
the gauge fields, 
\ba
\label{paritypreserving}
\mathcal{L}_0 = && R
- \frac{1}{2} G_{IJ}(\vartheta^K)\pd_a \vartheta^I \pd^a \vartheta^J - V(\vartheta^K) \cr
&& ~~
- \ell^2 Z_{PQ}(\vartheta^K) F^P_{ab}F^{Qab} ,
\ea
and $\mathcal{L}_{\rm CS}$ contains the axionic couplings,
\be
\label{CSterms}
\mathcal{L}_{\rm CS}
= - C_{PQ}(\vartheta^K) \SF^{Pab} F^Q_{ab}
  - {1 \ov 4} C (\vartheta^K) \SR R  . 
\ee
$\SF$ denotes the dual of $F$ and similarly for $\SR$. The parameter $\kappa$ is related to the bulk Newton constant $G_4$ as $\kappa^2 = 8 \pi G_4$ and  
$\ell$ is the radius of the anti-de Sitter (AdS) space.

We consider a most general bulk solution consistent with translational and rotational 
symmetries along boundary directions, 
\be
  \label{bmetric}
 ds^2 =  \frac{\ell^2}{z^2} \lb -f(z) dt^2 +h(z)dz^2 + \lb dx^i \rb^2 \rb \ 
\ee
with $z=0$ at the boundary and a horizon at $z = z_0$. The scalar fields $\vt^I$ are functions of $z$ only and 
the only nonzero component of $A_a^P$ is $A_t^P$ which again depends only on $z$. 
We denote the temperature as $T$ and the chemical potential associated with the boundary conserved current dual to the bulk gauge 
field $A^P_a$ as $\mu^P$. Thus $A_t^P (z=0)=\mu^P$ and at the horizon regularity requires $A_t^P(z=z_0)=0$. 

The angular momentum density $\mathcal{J}$ computed in \cite{Liutwo} can be expressed as a sum of two terms, 
\be
\label{totalangular}
\mathcal{J} = \mathcal{J}_{\rm horizon} + \mathcal{J}_{\rm integral}.
\ee
The first term in the right-hand side depends only on bulk fields at the horizon ($z=z_0$),
\be
\label{angularhorizon}
      \mathcal{J}_{\rm horizon} = -\frac{2\ell^2}{\kappa^2} \lsb
C_{PQ}\mu^P \mu^Q + 2\pi^2  C T^2 \rsb\Big|_{z=z_0}, 
\ee
and the second term is an integral from the horizon to the boundary,
\ba
\label{angularintegral}
&& \mathcal{J}_{{\rm integral}} \cr
&&= \frac{2\ell^2}{\kappa^2}\int_0^{z_0} dz \lsb C_{PQ}' (A_t^P - \mu^P)(A_t^Q - \mu^Q) 
+  \frac{C' f'^2}{8fh} \rsb, \nonumber \\
\ea
where $'$ indicates the derivative with respect to the bulk coordinate $z$. Of course by adding a total derivative term to the integrand of the bulk integral~\eqref{angularintegral} one can change the horizon 
piece and may also generate a boundary contribution. Other than that the split in~\eqref{angularhorizon}--\eqref{angularintegral} appears most naturally in the calculation of~\cite{Liutwo}, there is a sense in which the 
split is canonical as follows.  In~\cite{Liuone}, we found that when the scalar fields $\vt^I$ are dual to  marginal perturbations on the boundary, the angular momentum can be expressed solely in terms of 
quantities at the horizon. The split~\eqref{angularhorizon}--\eqref{angularintegral} has the properties that 
in the marginal case the integral part $\mathcal{J}_{{\rm integral}}$ vanishes identically (as $\vt^I$ are $z$-independent in this case). Thus it appears meaningful to interpret~\eqref{angularhorizon} as contribution from the IR physics and~\eqref{angularintegral} as contributions from other scales.

The Hall viscosity for Einstein gravity coupled to a single scalar field with gravitational Chern-Simons 
coupling~\eqref{bcs} was first derived in~\cite{SaremiSon}, and explicit computations for some specific
models have been done in  \cite{Chen:2011fs} and \cite{Chen:2012ti}. 
It can be readily generalized to the most general Lagrangian~\eqref{totalLag}--\eqref{CSterms}, and 
remarkably the same formula still applies, which in our notation can be written as 
\be
\label{Hallformula}
\eta_H = \frac{\ell^2}{4\kappa^2} \frac{C'f'}{fh}\Big|_{z=z_0}.
\ee
In particular, the gauge Chern-Simons term~\eqref{acs} does not give a contribution~\cite{Jensen:2011xb}.
The reason is as follows. The Hall viscosity can be obtained from linear response of the tensor sector, for instance by turning on a time-dependent source  in $h_{xx} - h_{yy}$ and measuring the linear response in $h_{xy}$. Since the linearized equations of motion for the gauge fields decouple from the tensor modes, the gauge Chern-Simons term does not contribute to the Hall viscosity.

We note an intriguing connection between~\eqref{Hallformula} and the second term of~\eqref{angularintegral}. 
Denoting 
\be 
A = \frac{\ell^2}{4\kappa^2} \frac{C'f'}{fh} 
\ee
we can write~\eqref{Hallformula} as 
\be 
\eta_H = A |_{\rm horizon}
\ee
while the second term  of~\eqref{angularintegral} can be written as 
\be 
-  \int_0^1 df \, A 
\ee
where we have changed the integration variable to the red-shift factor $f$.\footnote{The change of variable is legitimate as the redshift factor $f(z)$ should be a monotonic function of $z$ from the IR/UV connection.}

\section{Holographic RG flows: breaking by a scalar source on the boundary}

The expressions~\eqref{totalangular}--\eqref{angularintegral} and~\eqref{Hallformula} are somewhat formal 
as they are expressed in terms of abstract bulk gravity solutions,  which do not always have immediate boundary interpretations. To gain intuition on their physical behavior it is instructive to examine the explicit values of 
these expressions in some simple models. 

In this section we consider a class of holographic RG flows at a finite temperature/chemical potential, where 
the parity and time-reversal symmetries are broken by introducing a source for the scalar field $\vartheta$.
This corresponds to turning on a perturbation on the boundary by the operator dual to $\vartheta$.  
In next section we consider a class of models where the symmetries are broken by a dilaton coupling.

\subsection{Outline of the model}   
  
The simplest model with both non-vanishing angular momentum and Hall viscosity 
consists of one scalar field $\vartheta$ with the gravitational Chern-Simons coupling, 
\be
\label{Lagrsec2}
\mathcal{L}_{1}=\frac{1}{2\kappa^2} \sqrt{-g}\lsb R
- \frac{1}{2} \pd_a \vartheta \pd^a \vartheta - V(\vartheta) - {\alpha_{CS} \ov 4} \ell^2 \vartheta \SR R \rsb
\ee
where $\alpha_{CS}$ is a constant. In order for $\vt$ to have a nontrivial radial profile 
as is required for the non-vanishing of Hall viscosity~\eqref{Hallformula},  
we consider a potential $V (\vt)$ for which  $\vt$ is dual to a relevant boundary operator $\sO$. Recall that the mass $m$ of a scalar field is related to the conformal dimension $\Delta$ of the dual operator on the boundary by $m^2 \ell^2 = \Delta(\Delta-3)$, and near the AdS boundary we should have 
\be 
\vt (z) \to \vt_0 z^{\De_-} + v z^{\De_+}  \qquad z \to 0
\ee
where 
\be
  \Delta_\pm = \frac{3 \pm \sqrt{4m^2 \ell^2 + 9}}{2}, \qquad \De_+ = \De .
\ee
We turn on a uniform non-normalizable mode $\vartheta_0$, which corresponds to turning on a relevant perturbation $ \vt_0 \int d^3 x \, \sO$ in 
the boundary theory.  Since $\sO$ is odd under parity and time reversal, these symmetries are broken explicitly. At zero temperature, the system is described by a Lorentz invariant vacuum flow, and of course both angular momentum and Hall viscosity are zero. A nonzero angular momentum density and Hall viscosity can be  generated by putting the system at a finite temperature $T$ which then cuts off the flow at scale $T$. The bulk gravity solution (at a finite $T$) is described by a black brane of the form~\eqref{bmetric} with a nontrivial scalar profile. 

For~\eqref{Lagrsec2}, equations~\eqref{angularhorizon}--\eqref{Hallformula} become 
\ba
\label{currentJ}
\mathcal{J}_\mrm{horizon} &=& - \frac{4\pi^2 \aCS \ell^2}{\kappa^2} T^2 \vartheta (z_0) \\
\label{currentJ2}
\mathcal{J}_\mrm{integral}&=& 
\frac{\aCS \ell^2}{4\kappa^2} \intop_0^{z_0} dz \frac{f'^2}{fh} \vartheta' \\
\label{currentJ3}
\eta_H & = &  \frac{\aCS \ell^2}{4\kappa^2} \frac{\vt' (z_0) f' (z_0)}{f (z_0) h (z_0) }.
\label{hall1}
\ea
The bulk gravity solution depends on the specific form of the potential $V (\vt)$, and 
as we will see explicitly below so do~\eqref{currentJ}--\eqref{hall1}. From the boundary perspective different $V (\vt)$ correspond to 
different flows, which implies that the behavior of the angular momentum and Hall viscosity in general depends on specific flows. 

The gravity description suggests, however, that in the limit  $\vt_0 \to 0$, the behavior of these quantities should be ``universal,'' i.e. independent of the specific form of $V (\vt)$.
More explicitly, in this limit, throughout the flow, i.e. from the boundary to the horizon, $\vt$ is small. 
At leading order the nonlinear terms in $V(\vt)$ can be neglected, and we can simply replace it 
by the Gaussian potential 
\be
\label{gaussiantwo}
V(\vartheta) = - \frac{6}{\ell^2} + \frac{m^2 \vartheta^2}{2} .
\ee
Note that this argument for universality works not only for~\eqref{Lagrsec2}, but also 
for the general models of~\eqref{paritypreserving} and~\eqref{CSterms} (for the moment let us assume the 
gauge fields are not turned on). Note that since $\vt_0$ has dimension $\De_- = d - \De$, the appropriate dimensionless 
parameter is 
\be \label{smae} 
\ep \equiv \vt_0 T^{- \De_-} \to 0 .
\ee 

This universal limit also has a natural interpretation from the boundary side; when $\ep$ is small, 
we expect that the effect of parity and time reversal breaking can be captured by conformal perturbation theory 
near the UV fixed point, i.e. the angular momentum and Hall viscosity may be controlled by properties of $\sO$ at the UV fixed point, and not by details of the RG trajectories. 
It would be interesting to calculate angular momentum and Hall viscosity using conformal perturbation theory which we defer to later work.

\subsection{Leading results in the small $\vt_0$ limit}

In the small $\vt_0$ limit, to leading order we can approximate 
the potential $V(\vt)$ by~\eqref{gaussiantwo} and neglect the backreaction of scalar field 
to the background geometry. Thus we use the standard black brane metric with
\be
\label{metricSCH}
f = {1 \ov h} = 1 - {z^3 \ov z_0^3}
\ee
and treat the scalar field as a probe.  Since the scalar equation from~\eqref{gaussiantwo} is linear and in this limit the metric is independent of $\vt$, it follows from~\eqref{currentJ}--\eqref{hall1} that, at leading order, 
these expressions must be linear in $\vt_0$. This is of course consistent with the expectation from conformal perturbation theory as $\vt_0$ is a relevant boundary coupling. 
Given that both $\sJ$ and $\eta_H$ have dimension $2$, and the dimensionless 
expansion parameter is $\ep$~\eqref{smae}, we conclude on dimensional grounds that at leading order 
\ba \label{j1}
\sJ & \to  & c_J  \ep T^2 = c_J \vt_0 T^{2 - \De_-}   , \\
\eta_H & \to &  c_\eta \ep T^2 = c_\eta \vt_0 T^{2 - \De_-}  
\label{e1}
\ea
where $c_J$ and $c_\eta$ are some constants, and $c_J$ can further be separated into 
\be \label{eke}
c_J = c_{\rm horizon} + c_{\rm integral}  .
\ee
Since $m^2$ (or dimension $\De$) is the only parameter in
the Gaussian limit~\eqref{gaussiantwo}, $c_{\rm horizon}, c_{\rm integral}$ and 
$c_\eta$ are functions of $\De$ only.  These functions can be worked out analytically (see Appendix~\ref{app:A}
for details and their explicit expressions). Although they are all very complicated functions of $\De$, it turns out the ratio 
of $\sJ_{\rm horizon}$ and $\eta_H$ (both of which only receive contributions at the horizon) 
is remarkably simple, given by
\be
\label{Jhornice}
\frac{J_\mrm{horizon}}{\eta_H} = -\frac{9}{m^2\ell^2} =  {9 \ov \De (3-\De)} .
\ee
Note that the above expression diverges in the marginal limit $m^2 \to 0$ where $\vt$ becomes 
a $z$-independent constant so that $\eta_H$ vanishes. The simplicity of the ratio is intriguing and suggests a possible common physical origin for both quantities.

The ratio of the integral contribution $\sJ_{\rm integral}$ to $\eta_H$ is a rather complicated function 
of $m^2$ (see Appendix~\ref{app:A}), but can be expanded around $m^2 =0$ as 
\be \label{eme}
\frac{\sJ_\mrm{integral}}{\eta_H} = -\frac{3}{4} - 0.0383997 m^2 \ell^2 + O(m^4) .
\ee
The ratio has a finite $m^2 \to 0$ limit, as by design $\sJ_\mrm{integral}$ also 
approaches zero in the marginal limit. Note that due to coefficient of the second term
being rather small, and since $m^2 \in [-9/4, 0]$, equation~\eqref{eme} in fact gives a good global fit for the whole range\footnote{We take $m^2 \geq -9/4$ so that the Breitenlohner-Freedman bound is satisfied and $m^2 \leq 0$ so that the scalar field is either relevant or marginal. For $m^2 > 0$ the dual operator is irrelevant in which case the system requires a UV completion.} of $m^2$.

Combining with~\eqref{Jhornice}, we thus find that 
\ba
\frac{\mathcal{J}}{\eta_H} &=& - \frac{9}{m^2} - \frac{3}{4} + \mathcal{O}(m^2) \cr
&=& \frac{3}{\Delta_-} + \frac{1}{4} + \mathcal{O}(\Delta_-) 
\ea
for $\ep = \vartheta_0 T^{-\Delta_-}\rightarrow 0$.

\subsection{Generic $\vt_0$: numerical results} 

Away from the regime of small symmetry breaking, the results will depend on the explicit form of $V (\vt)$.
We now consider two classes of examples for illustration. 
As a first example, we consider the quadratic potential given by~\eqref{gaussiantwo}, but now
treated as a full toy potential. The other class we consider was introduced in \cite{Gao,Gao2} (the same potential 
also arose from a superpotential in the faked supergravity construction~\cite{Elvang:2007ba}, see also \cite{Garfinkle}), which we refer to as the Gao-Zhang potential after the authors of the paper, 
\ba
\label{gao-zhangpot}
V(\vartheta) &=& -\frac{2}{\ell^2}\frac{1}{(1+\alpha^2)^2}\big[\alpha^2\left(3\alpha^2-1\right)e^{-\frac{\vartheta}{\alpha}}
\nn\\
&& +\left(3-\alpha^2\right)e^{\alpha\vartheta}
+8\alpha^2e^{\frac{\alpha^2-1}{2\alpha}\vartheta}\big].
\ea
We will essentially use it as a proxy to a generic potential parameterized by some constant $\alpha$.
It should be noted that the quadratic part of the Gao-Zhang potential around $\vartheta=0$ always has $m^2=-2$.
We therefore can only compare its results with those obtained by the Gaussian potential with the same $m^2 = -2$. 

\begin{figure}
\centering
\includegraphics[width=8.5cm,clip=true]{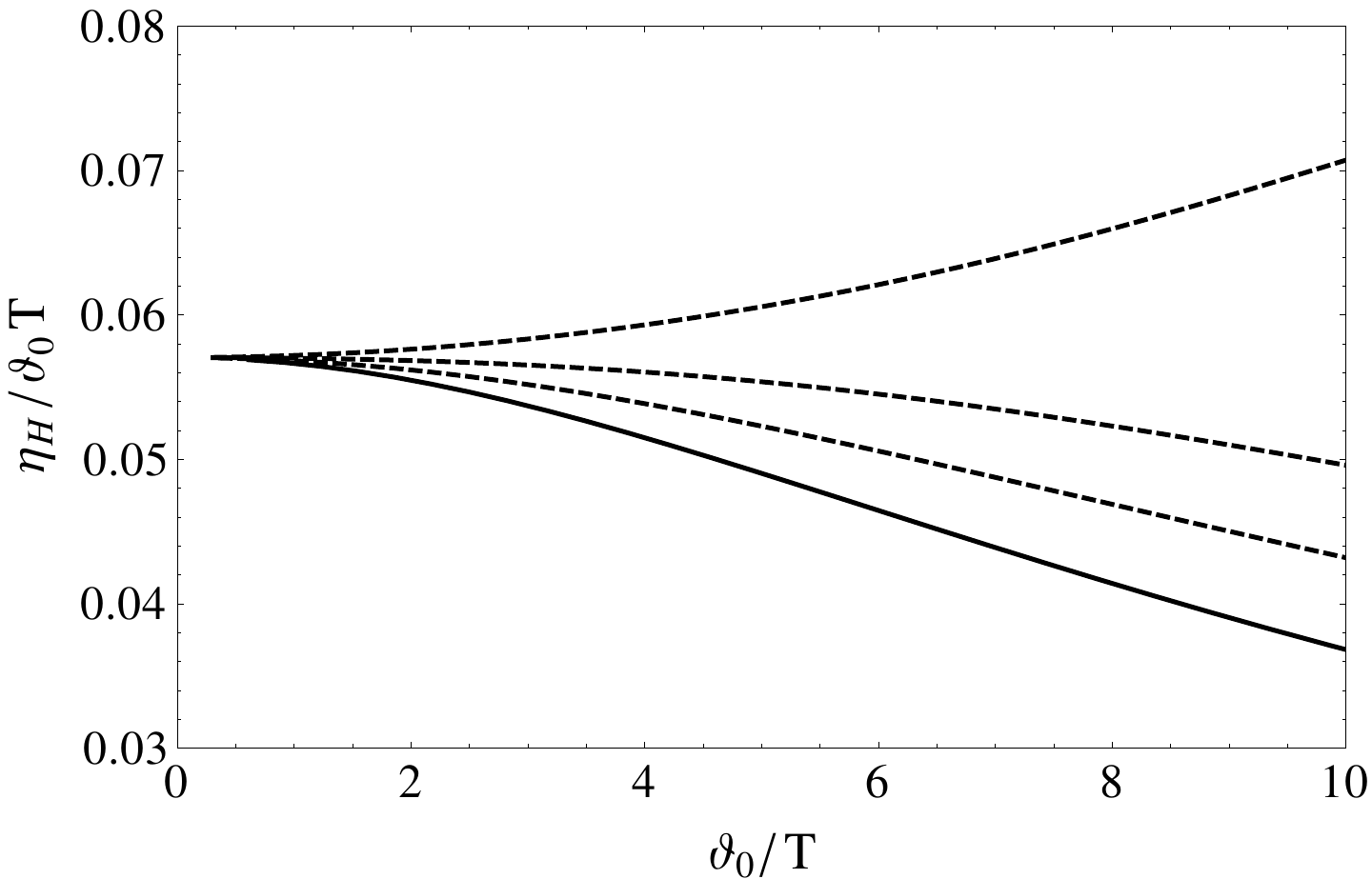}
\caption{(Color online) Hall viscosity for the Gao-Zhang potential \eqref{gao-zhangpot} with $\alpha=1.1$, $1.5$, $\sqrt{3}$ (dashed lines) and for a quadratic potential with $m^2=-2$ (solid line).}
\label{fig:msqm2_etaH}
\end{figure}

\begin{figure}
\centering
\includegraphics[width=8.5cm,clip=true]{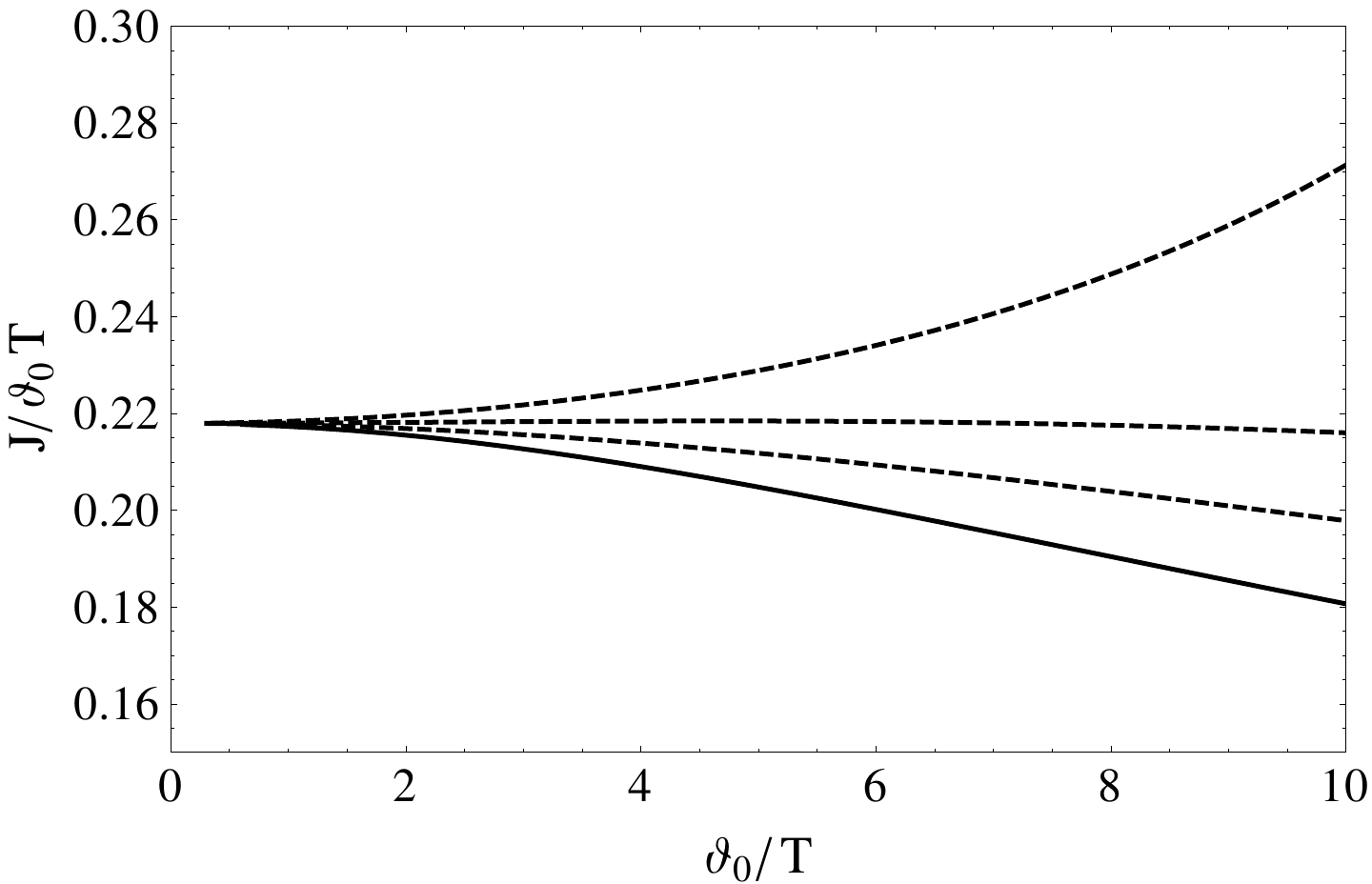}
\caption{(Color online) Angular momentum density for the Gao-Zhang potential \eqref{gao-zhangpot} with $\alpha=1.1$, $1.5$, $\sqrt{3}$ (dashed lines) and for a quadratic potential with $m^2=-2$ (solid line).}
\label{fig:msqm2_L}
\end{figure}

\begin{figure}
\centering
\includegraphics[width=8.5cm,clip=true]{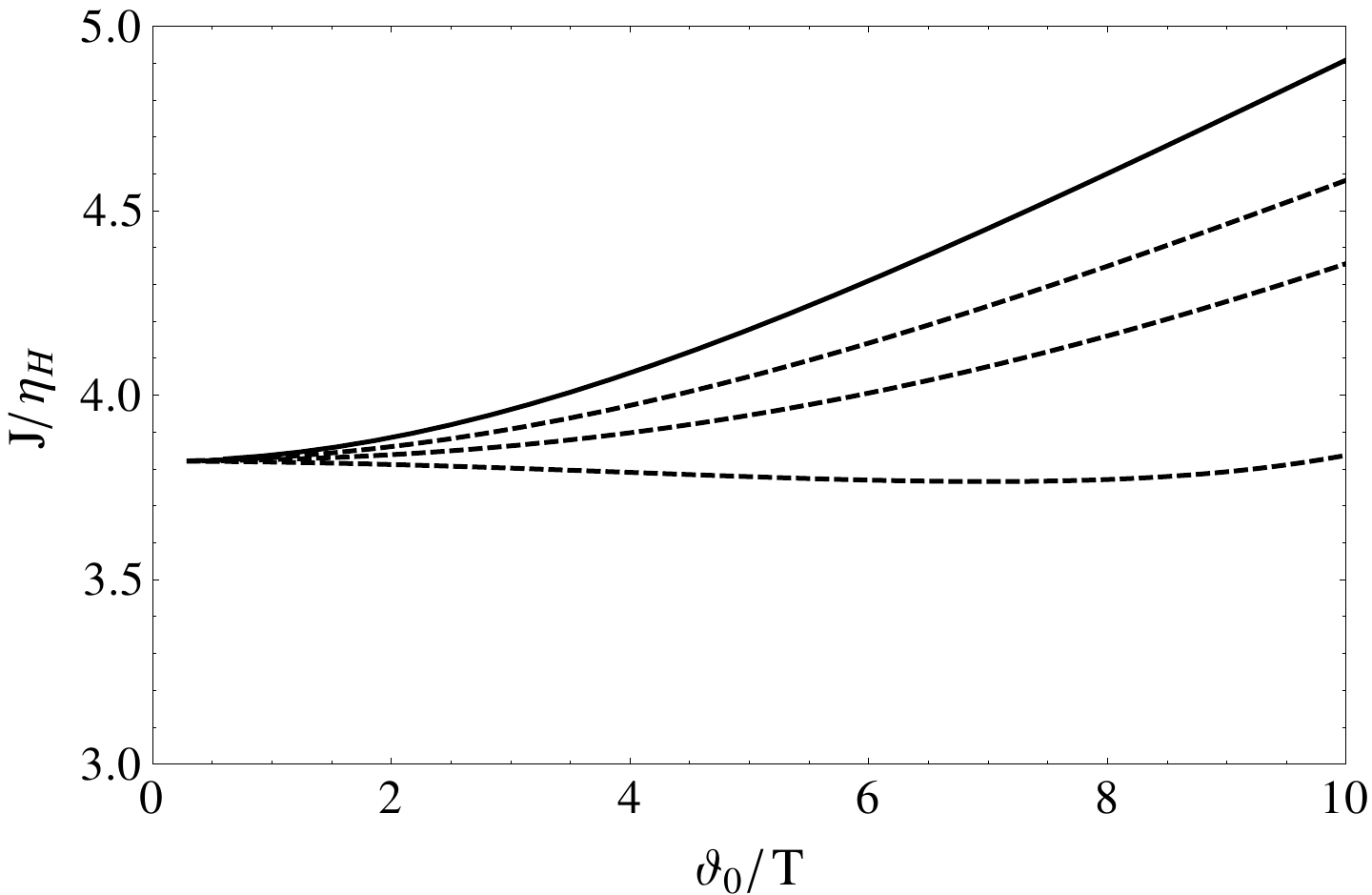}
\caption{(Color online) Ratio of angular momentum density and Hall viscosity for for the Gao-Zhang potential \eqref{gao-zhangpot} with $\alpha=1.1$, $1.5$, $\sqrt{3}$ (dashed lines) and for a quadratic potential with $m^2=-2$ (solid line).}
\label{fig:msqm2_R}
\end{figure}

For general $\vt_0$, the backreaction from the flow of $\vt$ to the metric can no longer be ignored.  
The gravity solution and equations~\eqref{currentJ}--\eqref{hall1} can now only be obtained numerically.  
Figs.~\ref{fig:msqm2_etaH} -- \ref{fig:msqm2_R} show the Hall viscosity $\eta_H$, angular momentum 
density $\mathcal{J}$ and their ratio for the two potentials. In these figures, all the curves corresponding to different potentials converge for $\vartheta_0 \ll T$, and approach finite values if we normalize them by 
$\vartheta_0 T$.  Since for $m^2 =-2$ we have $2 - \De_- = 1$, this confirms~\eqref{j1} -- \eqref{e1}.  
In particular, the numerical values of $c_{\rm horizon}, c_{\rm integral}$ and $c_\eta$, including the ratios~\eqref{Jhornice} -- \eqref{eme}, agree perfectly with those obtained from the analytic expressions of Appendix~\ref{app:A} for $m^2 =-2$.

\begin{figure}
\centering
\includegraphics[width=8.5cm,clip=true]{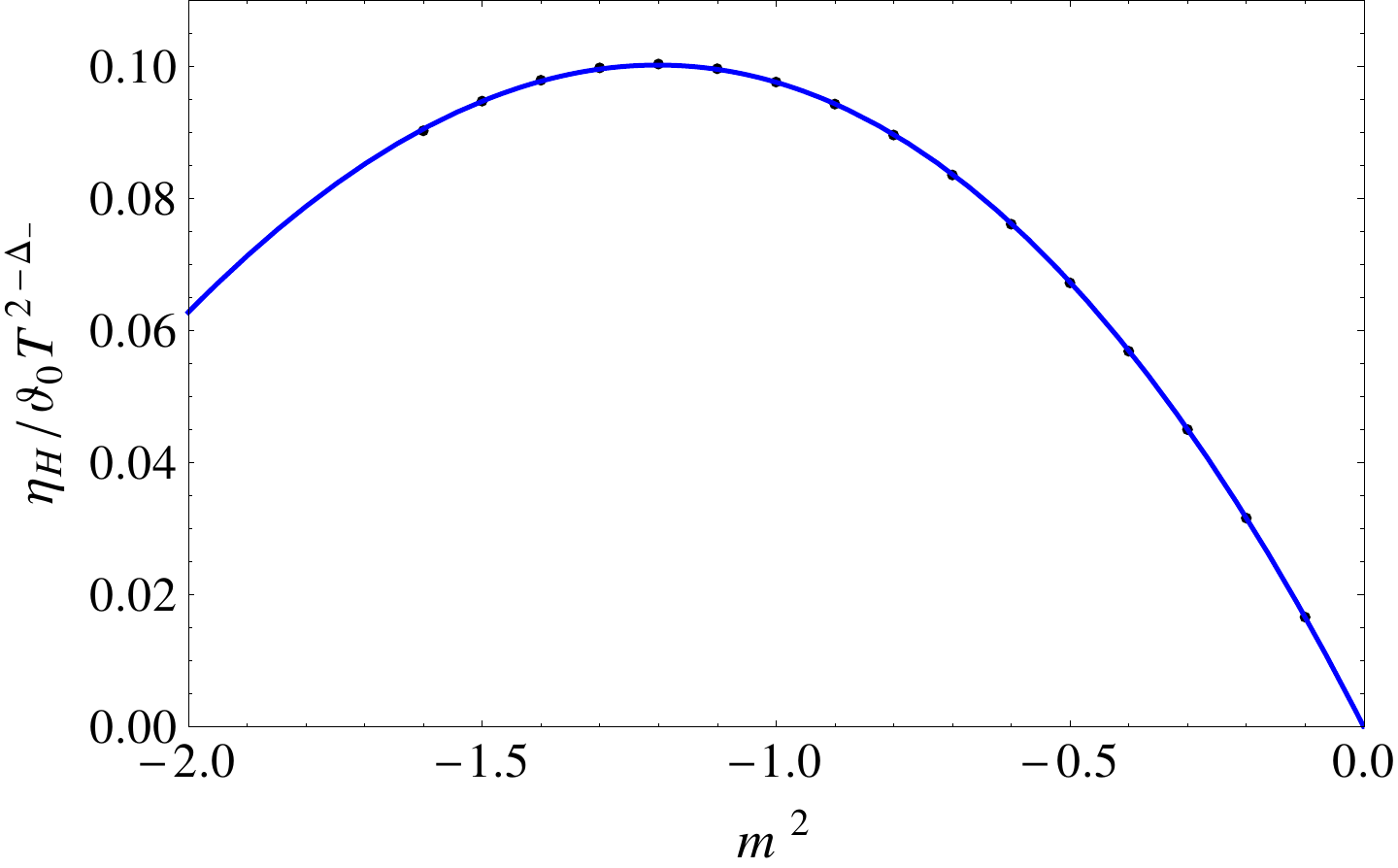}
\caption{(Color online) $\eta_H/\vartheta_0T^{2-\Delta_-}$ as a function of $m^2$ for $\vartheta_0/T^{\Delta_-}=1/6$.}
\label{fig:massanalysis_etaH}
\end{figure}

\begin{figure}
\centering
\includegraphics[width=8.5cm,clip=true]{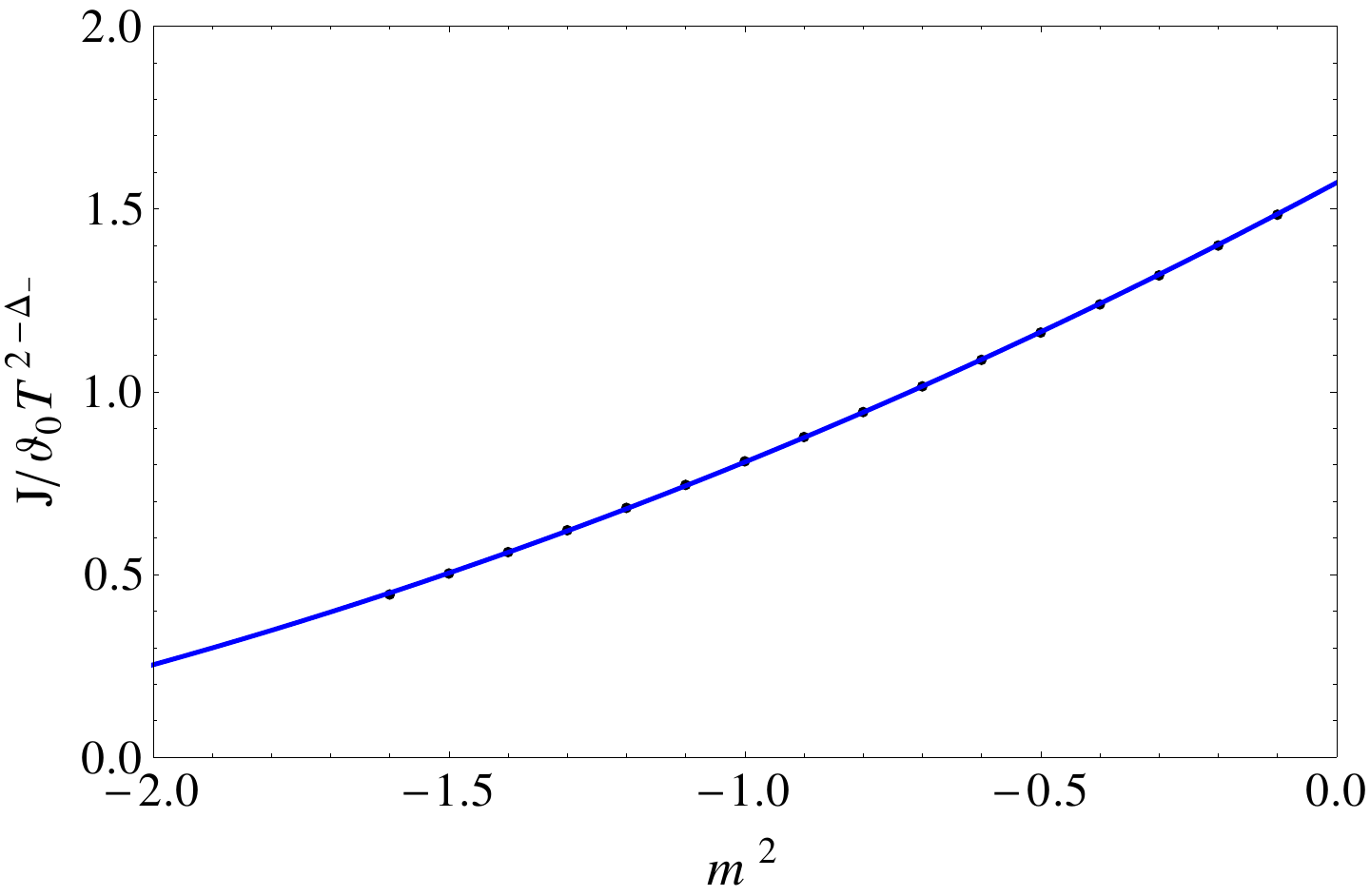}
\caption{(Color online) $\mathcal{J}/\vartheta_0T^{2-\Delta_-}$ as a function of $m^2$ for $\vartheta_0/T^{\Delta_-}=1/6$.}
\label{fig:massanalysis_L}
\end{figure}

\begin{figure}
\centering
\includegraphics[width=8.5cm,clip=true]{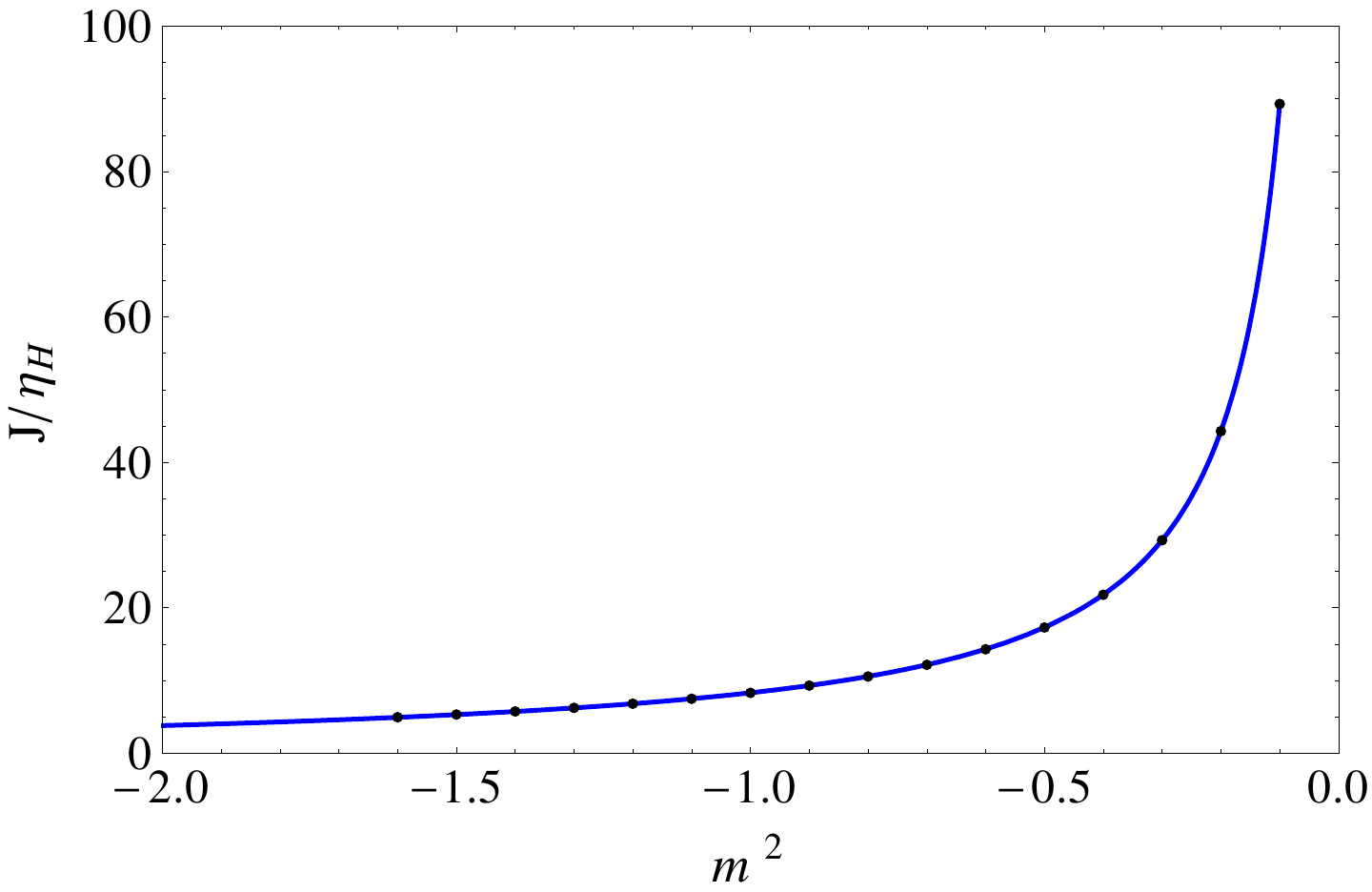}
\caption{(Color online) Angular momentum density to Hall viscosity ratio as a function of $m^2$ for $\vartheta_0/T^{\Delta_-}=1/6$.}
\label{fig:massanalysis_R}
\end{figure}

For arbitrary $m^2$, numerical results obtained for general $\vt_0$ using the Gaussian potential~\eqref{gaussiantwo} also agree very well with~\eqref{j1} -- \eqref{eme} obtained in the probe limit, which serves as a good consistency check for both calculations. 
In Figs.~\ref{fig:massanalysis_etaH} and \ref{fig:massanalysis_L}
we show $\eta_H$ and $\mathcal{J}$ as functions of $m^2$ obtained  at a fixed $\vartheta_0/T^{\Delta_-}=1/6$ 
that is sufficiently small to be in the plateau regime where $\eta_H/\vartheta_0 T^{2-\Delta_-}$ and $\mathcal{J}/\vartheta_0 T^{2-\Delta_-}$ 
are almost constant.
The ratio $\mathcal{J}/\eta_H$ is displayed in Fig. \ref{fig:massanalysis_R}, and we found it fitted well by the hyperbola
\be
\label{Rfit}
\frac{\mathcal{J}}{\eta_H} = - \frac{9}{m^2} + b .
\ee 
The coefficient $-9$ for the $1/m^2$ is within $0.1\%$ for the entire plateau interval from $\vartheta_0/T^{\Delta_-}=1/12$ 
to $1/4$. The constant term $b$ depends slightly on $\vartheta_0/T^{\Delta_-}$ as numerical fits show 
$-0.63$ at $\vartheta_0/T^{\Delta_-}=1/4$ and $-0.73$ at $\vartheta_0/T^{\Delta_-}=1/12$.
These results indicate that as we approach the limit $\vartheta_0/T^{\Delta_-}\rightarrow 0$, $b$ decreases monotonically and asymptotes to $-3/4$, which is consistent with~\eqref{eme}. 
One can fit with terms with higher orders in $m^2$. For example, including an additional $c m^2$ 
gives $c$ which is again consistent with value before $m^2$ in~\eqref{eme}. 
Similarly, excellent agreement with~\eqref{Jhornice}--\eqref{eme} is found for $\mathcal{J}_\mrm{horizon}/\eta_H$ and $\mathcal{J}_\mrm{integral}/\eta_H$ separately.

\subsection{Nonzero chemical potential}

Let us now consider turning on a chemical potential in the holographic RG flows discussed earlier. 
For this purpose we add a Maxwell term to the action~\eqref{Lagrsec2}, i.e. the Lagrangian density
becomes 
\be 
\sL = \sL_1 - {\ell^2 \ov 2 \kappa^2} \int d^4 x \sqrt{-g} F_{\mu \nu} F^{\mu \nu} \ 
\ee
and turn on a nonzero chemical potential for the gauge field, i.e. $A_t (z=0) = \mu$. 
We  are again interested in the leading behavior of the angular momentum and Hall viscosity 
in the limit $\vt_0 \to 0$, where we can replace the potential $V (\vt)$ by its quadratic 
order~\eqref{gaussiantwo} and treat the scalar field evolution along radial direction as a probe. 
Now the background geometry is replaced by that of  a charged black brane with
\be
\label{metricRRN}
f = {1 \ov h} = 1 - \frac{z^3}{z_M^3} + \frac{z^4}{z_Q^4},
\ee
where $z_M$ and $z_Q$ are length scales in the AdS bulk corresponding to the field theory energy and charge densities. They can be expressed in terms of temperature $T$ and chemical potential $\mu$ as
\be
z_M = \lb\frac{3}{4}\rb^{1/3} \frac{\left(\sqrt{3 \mu ^2+4 \pi ^2 T^2}-2 \pi 
   T\right)^{2/3}}{\mu^{4/3} \left(\sqrt{3 \mu ^2+4 \pi ^2 T^2}-\pi  T\right)^{1/3}},
\ee
\be
z_Q = \sqrt{\frac{\sqrt{3 \mu ^2+4 \pi ^2 T^2}-2 \pi  T}{\mu ^3}}
\ee
and the horizon is located at
\be
z_0 = \frac{\sqrt{3 \mu ^2+4 \pi ^2T^2}-2 \pi  T}{\mu ^2}.
\ee
On dimensional grounds we again write $\eta_H$ and $\mathcal{J}$ in the form~\eqref{j1} -- \eqref{eke}, except that the various coefficients are now functions of $\mu/T$, i.e. $c_J = c_J (m^2, \mu/T)$ and 
$c_{\eta} = c_{\eta} (m^2, \mu/T)$. Again although $c_{\rm horizon}$ and $c_\eta$ are complicated expressions, their ratio is remarkably simple and given by
\ba 
\label{ne}
\frac{\sJ_\mrm{horizon}}{\eta_H} &=& -\frac{16 \pi ^2 T^2 z_0^2}{m^2 \ell^2 } \\
&=& -\frac{16 \pi ^2 T^2 \left(\sqrt{3\mu ^2+4 \pi ^2  T^2} - 2\pi  T\right)^2}{\mu^4 m^2 \ell^2} \nn
\ea
where in the second line we have expressed the horizon size $z_0$ in terms of $\mu$ and $T$. 
Equation~\eqref{ne} recovers~\eqref{Jhornice} when $\mu =0$, but in the limit $T \to 0$ with 
a fixed $\mu$ it behaves as 
\be 
\frac{\sJ_\mrm{horizon}}{\eta_H}  = -\frac{48 \pi ^2 T^2 }{m^2 \ell^2 \mu^2 } , \quad T \to 0 .
\ee
The numerical analysis suggests this happens because $J_\mrm{horizon} \propto T^2$ at
small $T$, whereas $\eta_H \propto \mu^2$. The numerical analysis also indicates $J\propto \mu^2$ in the $T\to0$ limit, so that the ratio $J/\eta_H$ tends to a constant.

 \subsection{Analytic Gao-Zhang Solutions}
\label{subsec:AGZS}

As our last example of holographic RG flow we consider a Lagrangian with a dilatonic coupling 
in the Maxwell term, i.e. we add the following term to the Lagrangian of~\eqref{Lagrsec2}
\be 
\sL = \sL_1 - {\ell^2 \ov 2 \kappa^2} \int d^4 x \sqrt{-g} \, e^{- \al \vt} F_{\mu \nu} F^{\mu \nu} \ 
\ee
and turn on a nonzero chemical potential for the gauge field. In this case, with 
the potential given by~\eqref{gao-zhangpot}, there is a family of analytic solutions~\cite{Gao}, given by 
\ba
\label{pmmetric}
ds^2 &=& - f(r) dt^2 + f(r)^{-1} dr^2 + h(r) \left( dx^i \right)^2,  \\
f(r) &=& \frac{r_A}{r}\left(1+\frac{r_B}{r}\right)^{-\frac{2}{1+\alpha^2}} \nn\\
&& \times \left[ \frac{r^3}{\ell^2r_A} 
\left(1+\frac{r_B}{r}\right)^{\frac{4}{1+\alpha^2}}- 1\right], \\
h(r) &=& \frac{r^2}{\ell^2}\left(1+\frac{r_B}{r}\right)^\frac{2}{1+\alpha^2}\,,
\label{metric}
\ea
in coordinates that are convenient for our purpose. The solutions are parametrized by $\alpha$, with $\alpha=0$ corresponding to the Reissner--Nordstr\"om brane with scalar field turned off. The only non-vanishing component of the gauge field associated with the Maxwell tensor is
\be
A_t = \frac{Q}{r_0+r_B} - \frac{Q}{r+r_B},
\label{gaugefield}
\ee
while the scalar field is
\be 
\vartheta(r) = -\frac{2\alpha}{1+\alpha^2} \log\left( 1 +\frac{r_B}{r}\right).
\label{scalarfield}
\ee
Here $r_0$ is the location of the horizon, which appears in the gauge field because we impose the boundary 
condition $A_t (r=r_0) =0$. Note that the solution does not depend on the Chern-Simons coupling constant $\aCS$ 
since $\SR R=0$ for any spherically symmetric metric.

The electric charge density $Q$ of the black brane is given by
\beq
Q^2 = \frac{1}{\ell^2}\frac{r_Ar_B}{1+\alpha^2}.
\eeq
The corresponding chemical potential can be read off as
\beq
\mu = A_t(r=\infty) =  \frac{1}{\ell(r_0+ r_B)}\sqrt{\frac{r_Ar_B}{1+ \alpha^2}}.
\label{chem}
\eeq
The Hawking temperature is given by
\begin{align}
 T &= \frac{1}{4\pi} \frac{\partial f}{\partial r}{\Big|}_{r=r_0} = \frac{3}{4\pi \ell} \sqrt{\frac{r_A}{r_0}}\left[
 1 - \frac{4r_B}{3(1+\alpha^2)(r_0 + r_B)}\right].
\label{temp}
\end{align}

We should think of $r_{A/B}$ as given by the chemical potential in Eq.~(\ref{chem}) and by the temperature in Eq.~(\ref{temp}). The horizon location $r_0$ is obtained by solving $f(r_0) = 0$, namely,
\be
\frac{r_0^3}{\ell^2r_A} \left(1+\frac{r_B}{r_0}\right)^{\frac{4}{1+\alpha^2}}=1.
\label{horizon}
\ee 
Equations (\ref{chem}), (\ref{temp}) and (\ref{horizon}) can be solved analytically and we can thus
express $r_{A/B}$ and $r_0$ in terms of algebraic functions of $\mu$ and $T$. For $\alpha^2 < 1/3$ there is only one solution for given values of $\mu$ and $T$,
\be
\label{twosolutions}
r_0 = \sqrt{1+\alpha^2}\ell^2\frac{\mu}{U}\lb 1 + U^2 \rb^\frac{\alpha^2-1}{\alpha^2+1}
\ee
where
\be
\label{upm}
U = \frac{\sqrt{1+\alpha^2}\lb\sqrt{4\pi^2T^2 + 3\lb1-3\alpha^2\rb\mu^2} -2\pi T \rb}{\lb 1-3\alpha^2 \rb\mu}.
\ee
For $\alpha^2\geq 1/3$ there are two solutions, corresponding to different values of the scalar field source $\vartheta_0$. For simplicity, however, in this paper we will focus on the $\alpha^2< 1/3$ case, as taking $\alpha^2\geq1/3$ does not offer further physical intuition.

Let's now examine the behavior of the scalar field near the
boundary. Expanding $\vartheta$ as $r \to \infty$, we obtain
\ba
  \vartheta &=& -\frac{2\alpha}{1+\alpha^2} \frac{r_B}{r} 
               + \frac{\alpha}{1+\alpha^2}\left(\frac{r_B}{r}\right)^2 + \cdots.
\ea
The CFT operator $\Phi$ dual to $\vartheta$ carries conformal dimension $1$ or $2$ depending on the boundary condition for $\vartheta$.
To be more specific, let us pick the boundary condition
so that the conformal dimension of $\Phi$ is $1$. If one wants the conformal dimension to be $2$, 
one can simply exchange the expectation value $\langle \Phi \rangle$ and the source $\vartheta$ in the discussion below. 

According to the standard AdS/CFT dictionary, the coefficient of $1/r^2$ should be interpreted as an expectation value $\langle \Phi \rangle$ of the CFT operator $\Phi$. The coefficient of the $1/r$ term 
is then interpreted as a source $\vartheta_0$ for $\Phi$. The nonvanishing $1/r$ term in the expansion of $\vartheta$ 
means that the dual CFT is deformed by turning on the relevant operator $\Phi$. The magnitude of the 
deformation, $\vartheta_0$, is proportional to $r_B$,
\be
\label{Misthis}
 \vartheta_0 = -\frac{2\alpha}{1+\alpha^2} r_B = -\frac{2\alpha}{\sqrt{1+\alpha^2}} \ell^2\mu U \lb 1 + U^2 \rb^\frac{\alpha^2-1}{\alpha^2+1},
\ee 
so that $\vartheta_0/T$ is a function of $\mu/T$ and $\alpha$. Since different deformations correspond to different CFTs, 
in the bulk a given value of $\vartheta_0/T$ is related to a fixed value of $\mu/T$. 

Even though in the Gao-Zhang setup the scalar source is not independent of $\mu$ and $T$, their model can be used to gain analytic intuition into the relation between Hall viscosity and angular momentum density. In particular, \cite{Hallviscostwo} has argued from the field theory side that there should exist a simple proportionality relation between the two, at least in certain gapped systems. This can be readily compared with the analytic model of Gao-Zhang, and we find that for this class of models the relation between the two quantities is more complicated.

Applying Eqns. \eqref{currentJ} -- \eqref{currentJ3} to the Gao-Zhang model we obtain for the Hall viscosity and gravitational Chern-Simons angular momentum density
\ba
\eta_H = \frac{\aCS }{4\kappa^2} h(r_0)f'(r_0)\vartheta'(r_0)
\ea
and
\ba
\label{angmomGZ}
\mathcal{J} &=&  \frac{\aCS}{4\kappa^2} \intop_{r_0}^\infty dr\ \vartheta(r) \pd_r \lsb h^2(r) \lb \pd_r \frac{f(r)}{h(r)} \rb^2 \rsb
\ea
with $f$, $h$ and $\vartheta$ given in Eqns. \eqref{pmmetric}--\eqref{scalarfield}. $\mathcal{J}_\mrm{horizon}$ is still specified by Eqn. \eqref{currentJ} with the scalar field evaluated at the horizon.

The Hall viscosity can further be written in terms of $T$, $r_0$, $r_A$ and $r_B$ as
\be
\eta_H =  \frac{2\pi\aCS \ell}{\kappa^2}\frac{\alpha}{1+\alpha^2}T\sqrt{\frac{r_A}{r_0}}\frac{r_B}{r_0+r_B},
\ee
which can then be recast in terms of $\mu$ and $T$ as
\be
\eta_H = \frac{2 \pi\aCS \ell^2 \alpha T \left(\sqrt{3 \left(1-3
   \alpha ^2\right) \mu ^2+4 \pi ^2 T^2} - 2 \pi  T \right)}{\left(1 - 3 \alpha^2\right) \kappa^2}.
\ee

The total angular momentum in Eqn. \eqref{angmomGZ} can also be integrated in closed form, but the result is long and unilluminating. In terms of $\mu$ and $T$ the horizon component of the angular momentum reads
\be
\mathcal{J}_\mrm{horizon} = \frac{8 \pi ^2 \aCS \ell^2 \alpha T^2}{\left(\alpha
   ^2+1\right) \kappa ^2} \ln \left(1 + U^2\right)
\ee
with $U$ defined in Eqn. \eqref{upm}.

To gain some physical intuition, we can expand the Hall viscosity, gravitational angular momentum and horizon component of the angular momentum in a series at small $\mu$ as
\be
\frac{\eta_H}{T^2} = \frac{\aCS\ell^2}{\kappa^2} \lsb \frac{3\alpha\mu^2}{2T^2} + \frac{9\alpha\left(3 \alpha ^2-1\right) \mu^4}{32\pi^2T^4} + \mathcal{O}\lb \frac{\mu^6}{T^6} \rb \rsb,
\ee
\be
\frac{\mathcal{J}}{T^2} = \frac{\aCS\ell^2}{\kappa^2} \lsb \frac{18 \alpha \mu ^2}{5T^2} + \frac{9 \alpha\left(57 \alpha^2-49\right)\mu ^4}{160\pi^2T^4} + \mathcal{O}\lb \frac{\mu^6}{T^6} \rb \rsb,
\ee
\be
\frac{\mathcal{J}_\mrm{horizon}}{T^2} = \frac{\aCS\ell^2}{\kappa^2} \lsb \frac{9 \alpha  \mu ^2}{2T^2}+\frac{27 \alpha  \left(9 \alpha ^2-7\right) \mu ^4}{64
   \pi ^2 T^4} + \mathcal{O}\lb \frac{\mu^6}{T^6} \rb \rsb.
\ee

Finally, when including an axionic term
\be
\mathcal{L}_{CS} = - \bCS \ell^2 \vartheta \SF^{ab}F_{ab}
\ee
the  angular momentum density for the Gao-Zhang solutions is
\be
\mathcal J = \frac{4\bCS\ell^2}{\kappa^2} \intop_{r_0}^\infty dr \lb A_t(r) - \mu \rb A_t'(r)\vartheta(r) .
\ee
This expression can be integrated in closed form but it is unilluminating. In the small $\mu$ limit it can be expanded as
\be
\frac{\mathcal{J}}{\mu^2} = \frac{\bCS\ell^2}{\kappa^2} \lsb \frac{3 \alpha \mu ^2}{2 \pi ^2 T^2} + \frac{9 \alpha  \left(33 \alpha ^2-31\right) \mu ^4}{256 \pi^4 T^4} + \mathcal{O}\lb\frac{\mu^6}{T^6}\rb \rsb.
\ee

\section{Holographic vev flow: breaking by the dilaton coupling}

We now consider a class of models in which the scalar $\vt$ is normalizable 
at the AdS boundary, but the parity and time-reversal symmetries are broken 
by the dilaton coupling. We consider the Lagrangian 
\be
\label{LIIB}
\mathcal{L}=\frac{1}{2\kappa^2} \sqrt{-g}\lsb R
- \frac{1}{2} \lb\pd \vartheta \rb^2 - V(\vartheta) - \ell^2 e^{-\alpha\vartheta} F^2 + \mathcal{L}_{\rm CS} \rsb
\ee
and put the system at finite chemical potential. In this setup the bulk gauge field which is needed for a nonzero chemical potential can drive a normalizable nontrivial scalar hair through the dilatonic coupling. The  parity and time-reversal symmetries are broken since the dilaton coupling $e^{-\alpha \vartheta}$ is not invariant under
$\vartheta \rightarrow -\vartheta$. 

Below, we will discuss 
the gauge Chern-Simons term and the gravitational Chern-Simons term separately. 
Since both the angular momentum and the Hall viscosity are linear in these couplings, we can simply add the two results to obtain the whole picture. 
Just as in the previous section, we use two types of potentials, the Gaussian potential \eqref{gaussiantwo} and 
the Gao-Zhang potential \eqref{gao-zhangpot}.

\subsubsection{Gauge Chern-Simons Coupling}
For a single scalar field, let us parametrize the gauge Chern-Simons term as
\be
\mathcal{L}_{CS} = - \bCS \ell^2 \vartheta \SF^{ab}F_{ab}.
\ee
This term generates angular momentum density but not Hall viscosity, which was 
first pointed out by \cite{Jensen:2011xb}. Specializing Eqns. \eqref{angularhorizon} and \eqref{angularintegral} to Lagrangian \eqref{LIIB} the angular momentum density is given by
\be
\label{Jaxis}
\mathcal{J}= - \frac{2 \bCS\ell^2}{\kappa^2}\mu^2\vartheta(z_0)
+ \frac{2 \bCS\ell^2}{\kappa^2} \intop^{z_0}_0 dz \lb A_t(z) - \mu \rb^2  \vartheta'(z)
\ee
while the gravitational Chern-Simons contribution is the same as in Eqns. \eqref{currentJ}--\eqref{currentJ2}.

\begin{figure}
\centering
\includegraphics[width=8.5cm,clip=true]{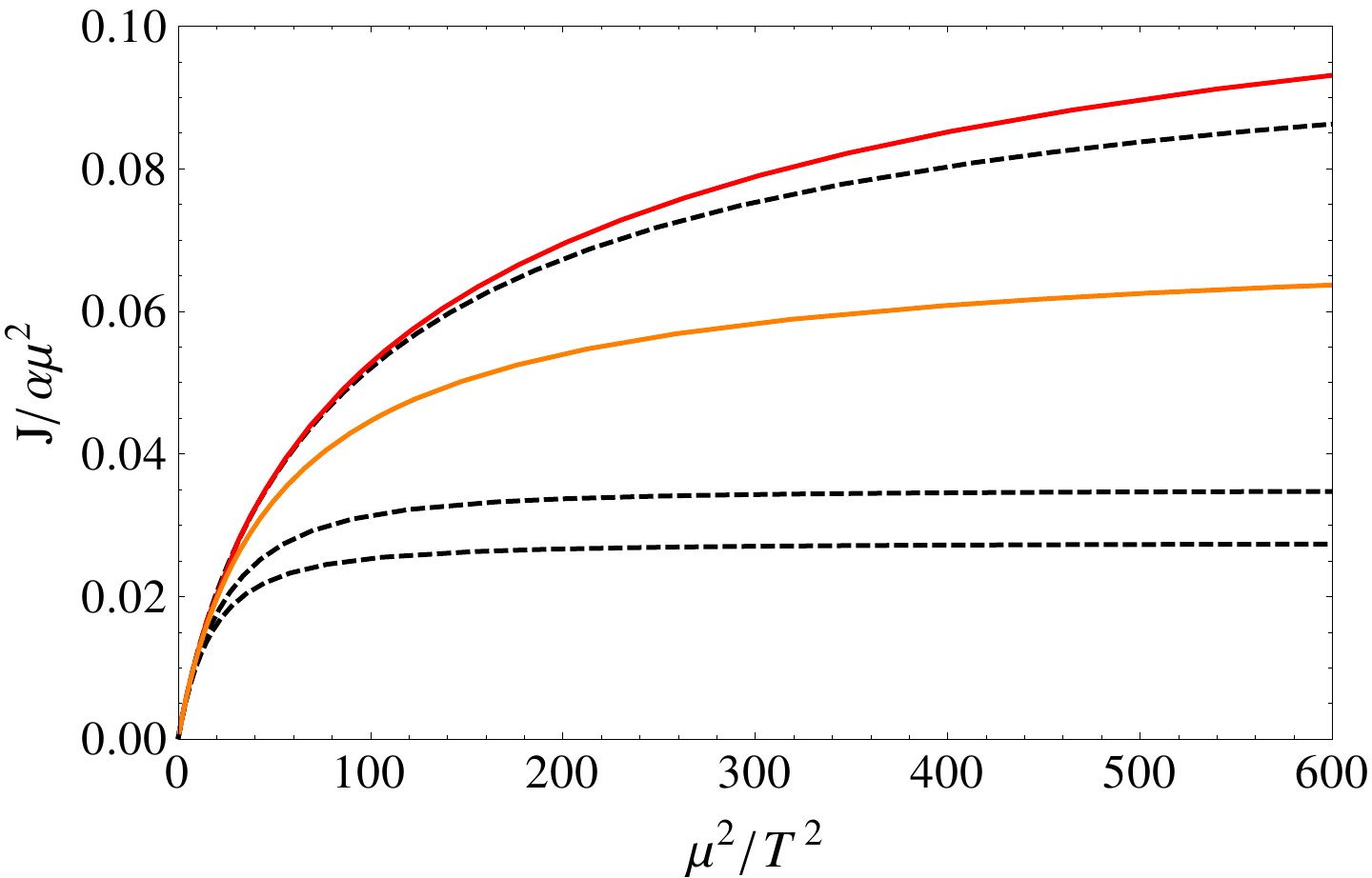}
\caption{(Color online) The angular momentum density generated by the gauge Chern-Simons coupling as a function of $\mu^2/T^2$ for the Gaussian potential $m^2=-2$ and dilatonic coupling $\alpha=0.5$ (red line) and $0.9$ (orange line) and for Gao-Zhang potentials with $\alpha=0.5$, $1.5$ and $\sqrt{3}$ (dashed lines).}
\label{fig:ax_J_potentials}
\end{figure}

Fig. \eqref{fig:ax_J_potentials} shows the angular momentum density as a function of $\mu^2/T^2$ for a Gaussian potential with  $m^2=-2$ and Gao-Zhang potentials with various $\alpha$. 
The vertical axis is normalized by the dilatonic coupling $\alpha$. 
We note that in the small $\mu/T$ limit all curves scale as $\mathcal{J} \propto \alpha \mu^4/T^2$. 
This is to be expected since in this limit the scalar field can be expanded in a perturbative series with the first term 
proportional to $\alpha\mu^2/T^2$ (by parity and dimensional analysis) and the gauge fields in Eqn. \eqref{Jaxis} contribute 
a factor of $\mu^2$. 

\begin{figure}
\centering
\includegraphics[width=8.5cm,clip=true]{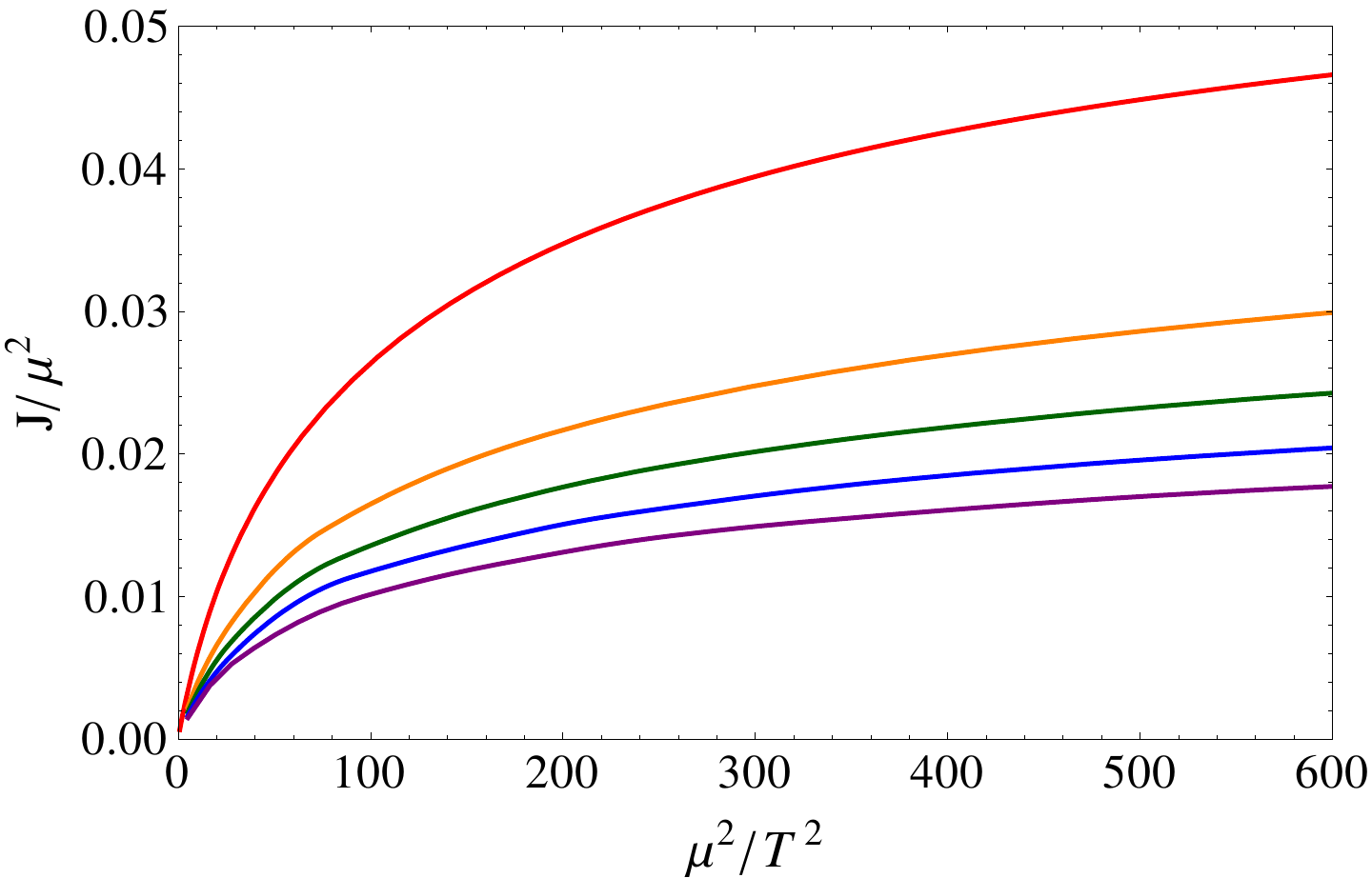}
\caption{(Color online) The angular momentum density generated by the gauge Chern-Simons coupling as a function of $\mu^2/T^2$ for dilatonic coupling $\alpha=0.5$ and Gaussian potentials with $m^2=-2$ (red), $1.4$ (orange), $-1$ (green), $-0.6$ (blue) and $-0.2$ (purple).}
\label{fig:ax_J_masses}
\end{figure}

Fig. \eqref{fig:ax_J_masses} shows the angular momentum as a function of $\mu^2/T^2$ for Gaussian potentials of various $m^2$. We note that although for all masses $\mathcal{J} \propto \alpha \mu^4/T^2$ in the small $\mu/T$ limit, the slope depends on $m^2$. 
We also remark that, for large $\mu/T$, the angular momentum density is proportional to $\mu^2$ for all the potentials we have investigated. Since the vertical axes of these figures are taken to be $\mathcal{J}/\mu^2$, this can be seen in some of the curves becoming horizontal for large $\mu^2/T^2$. We note that all curves flatten at large $\mu^2/T^2$, even though this is not apparent in the range displayed in the figures.
The asymptotic value of $\mathcal{J}/\mu^2$ depends on the dilatonic coupling, scalar field mass and on the details of the potential. 
It would be interesting to further explore this relation to better determine the type of models where it holds.

\subsubsection{Gravitational Chern-Simons Coupling}

We now turn out attention to the gravitational Chern-Simons coupling, adding
\be
\mathcal{L}_{CS} = - {\alpha_{CS} \ov 4} \ell^2 \vartheta \SR R
\ee
to Lagrangian \eqref{LIIB}. This will generate both Hall viscosity and angular momentum, according to Eqn. \eqref{currentJ3} for the Hall viscosity and to Eqns.~\eqref{currentJ}--\eqref{currentJ2} for the angular momentum density.

\begin{figure}
\centering
\includegraphics[width=8.5cm,clip=true]{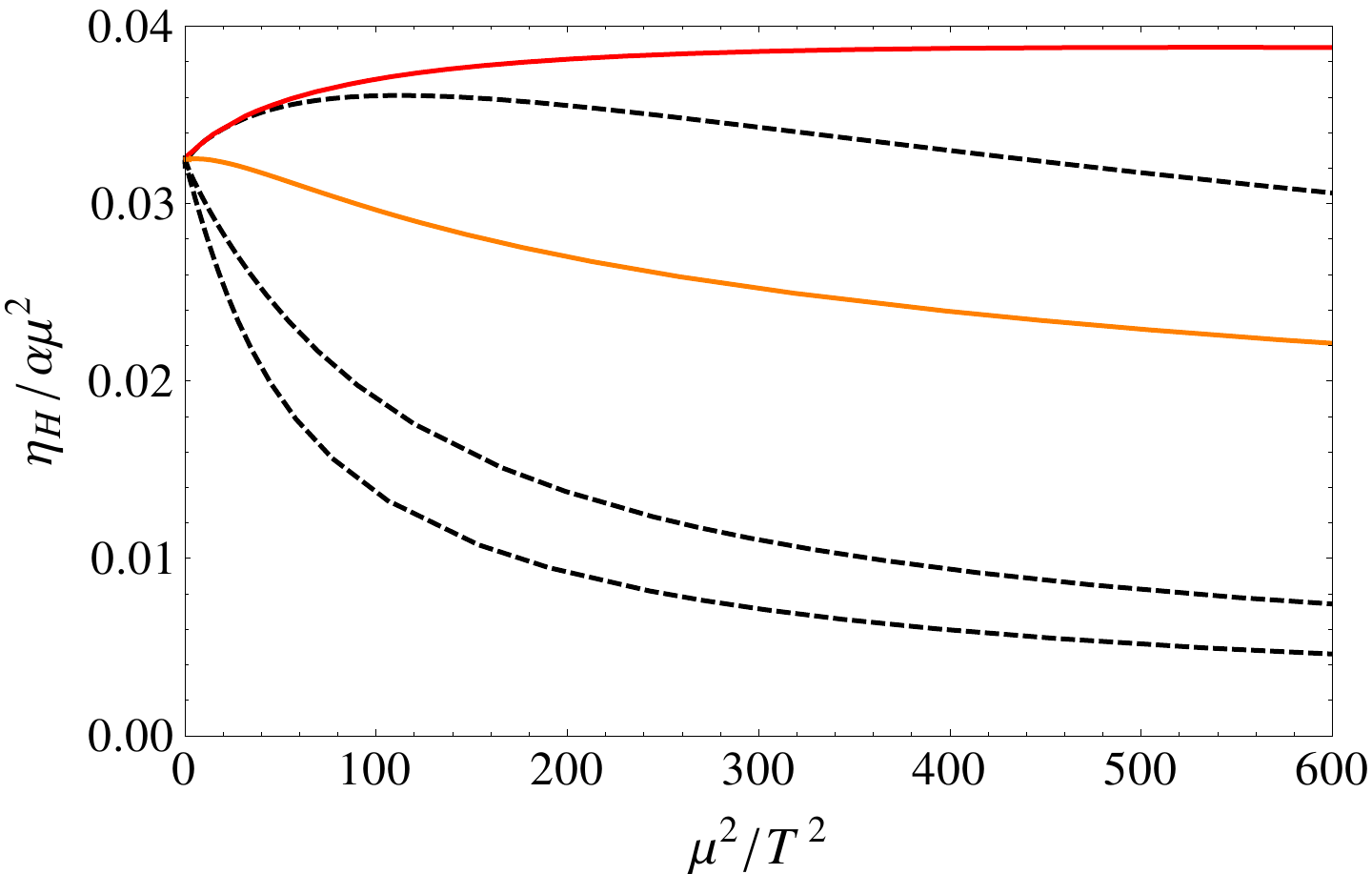}
\caption{(Color online) Hall viscosity as a function of $\mu^2/T^2$ for Gao-Zhang potentials with $\alpha=0.5$, $1.5$ and $\sqrt{3}$ (dashed lines) and for quadratic potentials with $m^2=-2$ and dilatonic coupling $\alpha=0.5$ (red line) and $0.9$ (orange line).}
\label{fig:IIB_gr_etaH_potentials}
\end{figure}

\begin{figure}
\centering
\includegraphics[width=8.5cm,clip=true]{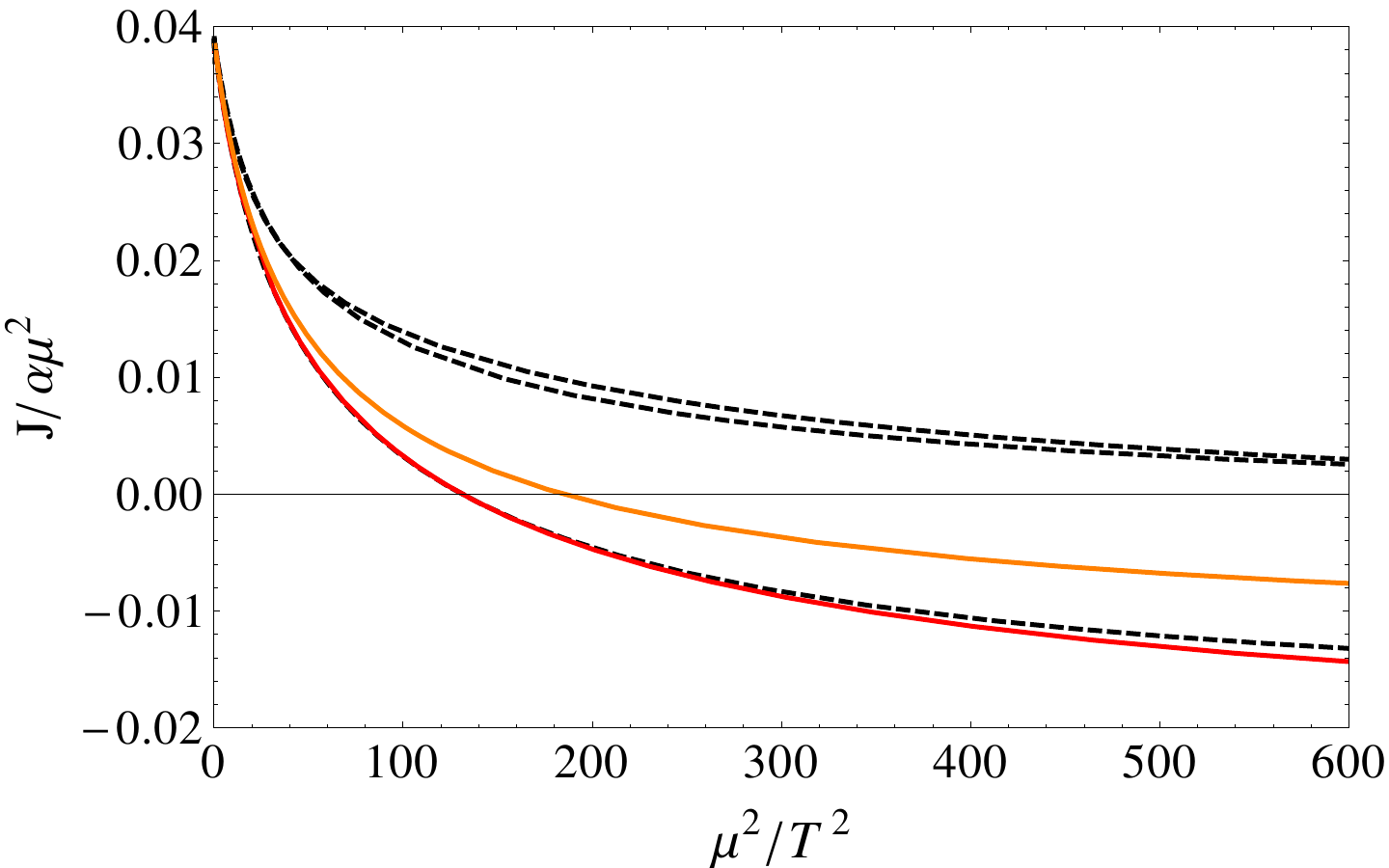}
\caption{(Color online) Gravitational angular momentum density as a function of $\mu^2/T^2$ for Gao-Zhang potentials with $\alpha=0.5$, $1.5$ and $\sqrt{3}$ (dashed lines) and for quadratic potentials with $m^2=-2$ and dilatonic coupling $\alpha=0.5$ (red line) and $0.9$ (orange line).}
\label{fig:IIB_gr_J_potentials}
\end{figure}

\begin{figure}
\centering
\includegraphics[width=8.5cm,clip=true]{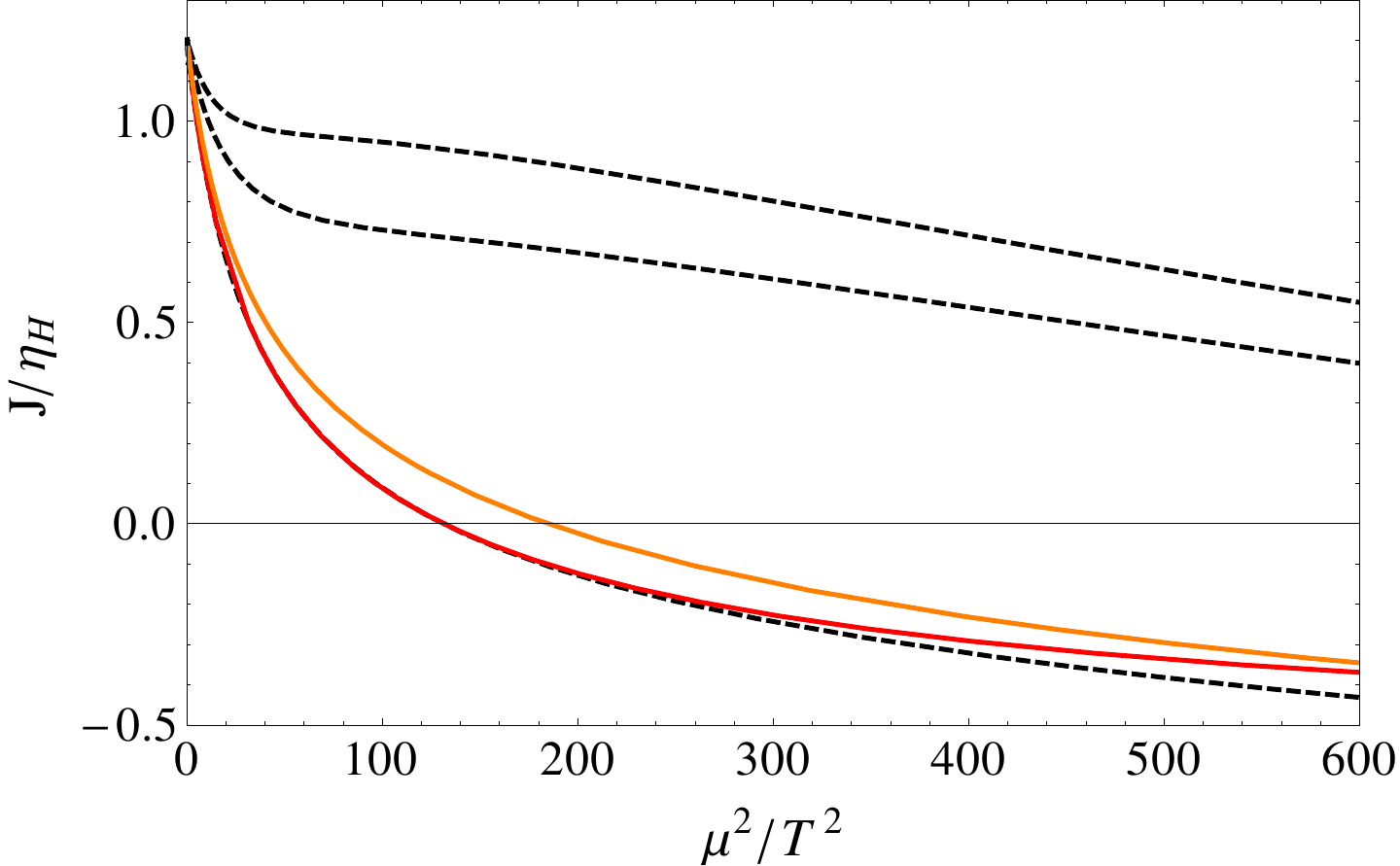}
\caption{(Color online) Gravitational angular momentum density to Hall viscosity ratio as a function of $\mu^2/T^2$ for Gao-Zhang potentials with $\alpha=0.5$, $1.5$ and $\sqrt{3}$ (dashed lines) and for quadratic potentials with $m^2=-2$ and dilatonic coupling $\alpha=0.5$ (red line) and $0.9$ (orange line).}
\label{fig:IIB_gr_R_potentials}
\end{figure}

Figs. \eqref{fig:IIB_gr_etaH_potentials} -- \eqref{fig:IIB_gr_R_potentials} show the Hall viscosity, the angular momentum density and their ratio as functions of $\mu^2/T^2$ for the Gaussian potential with $m^2=-2$ and for Gao-Zhang potential with various $\alpha$. For small $\mu/T$, all curves converge, and we have
\be
\eta_H = 0.032 \alpha \mu^2, \qquad \mathcal{J} = 0.039 \alpha \mu^2.
\ee

\begin{figure}
\centering
\includegraphics[width=8.5cm,clip=true]{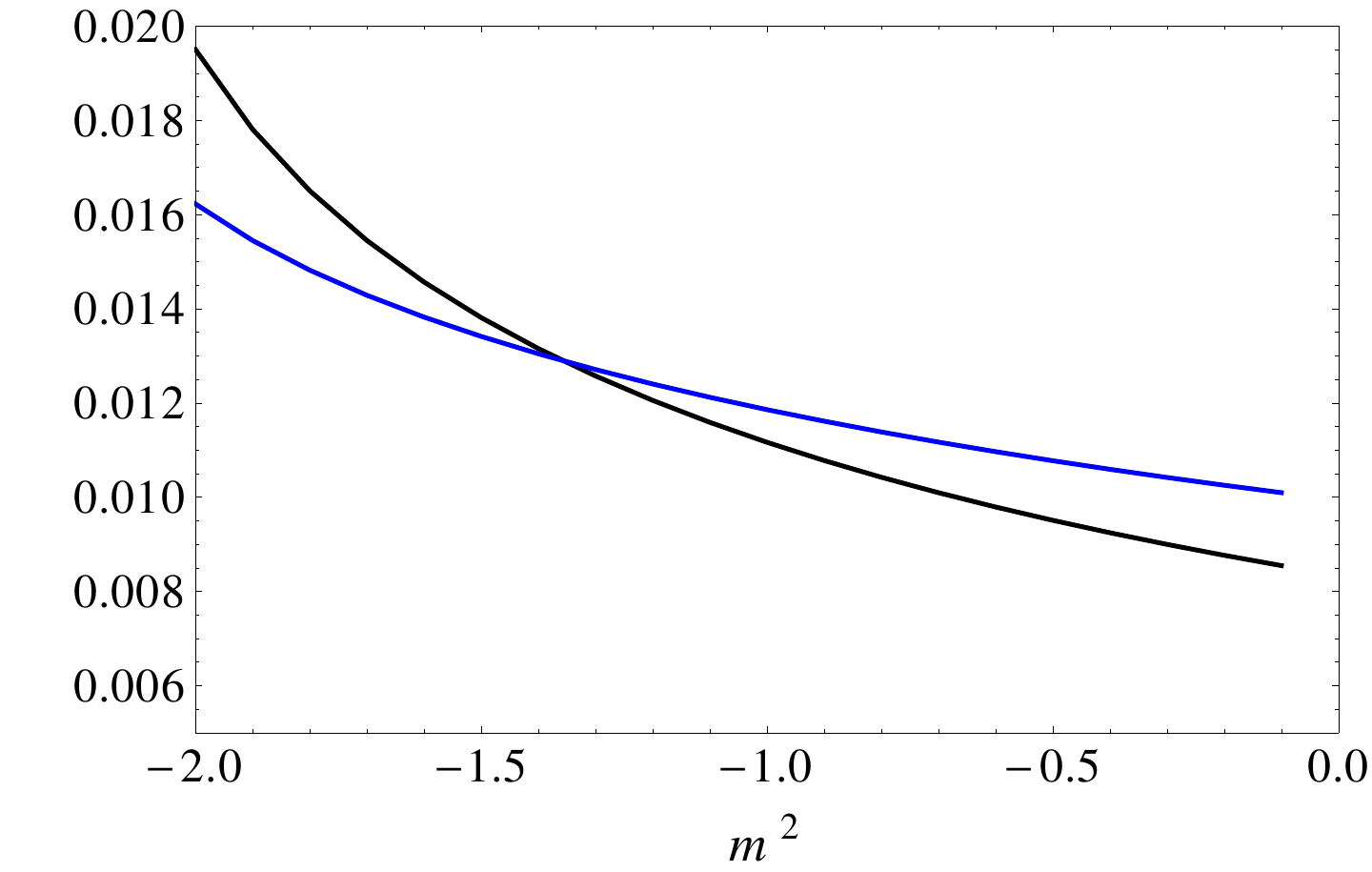}
\caption{(Color online) $\eta_H/\mu^2$ (blue) and $\mathcal{J}/\mu^2$ (black) as a function of $m^2$ for $\mu^2/T^2=0.1$.}
\label{fig:massanalysis_etaH_and_J}
\end{figure}

Fig. \eqref{fig:massanalysis_etaH_and_J} shows Hall viscosity and angular momentum density as functions 
of $m^2$ at $\mu^2/T^2=0.1$. This value of $\mu^2/T^2$ is 
sufficiently small to be in the plateau regime, and $\eta_H/\mu^2$ and $\mathcal{J}/\mu^2$ are $\mu/T$-independent. 
We note both $\eta_H$ and $\mathcal{J}$ are non-zero at $m^2\rightarrow 0$, and their values in this limit are
\be
\frac{\eta_H}{\mu^2} = 0.0099 + \mathcal{O}(m^2), \qquad \frac{\mathcal{J}}{\mu^2} = 0.0082 + \mathcal{O}(m^2).
\ee
The two numerical coefficients vary by less than $1\%$ as $\mu^2/T^2$ is increased from 0 to $\mu^2/T^2 \lesssim 0.6$. Both the horizon and the bulk terms contribute to the total angular momentum and are of the same order of magnitude, but have opposite signs.

The ratio $J/\eta_H$ is represented in Fig. \eqref{fig:massanalysis_R} for $\mu^2/T^2=0.1$. In the small $m$ limit it can be expanded as
\be
\frac{\mathcal{J}}{\eta_H} = 0.84 + \mathcal{O}(m^2)
\ee
where again the numerical coefficient varies by less than $1\%$ for $\mu^2/T^2 \lesssim 0.6$. For the horizon part of the angular momentum density we have
\be
\frac{\mathcal{J}_\mrm{horizon}}{\eta_H} = 1.33 + \mathcal{O}(m^2)
\ee
at $\mu^2/T^2=0.1$ and the numerical coefficient decreases by about $1\%$ at $\mu^2/T^2=0.6$. It is possible to better understand the ratio $\mathcal{J}_\mrm{horizon}/\eta_H$ by going to the probe approximation and using the scalar field equation to relate $\vartheta$ and $\vartheta'(z_0)$. Doing so gives
\be
\label{Ratiodilat}
\frac{\mathcal{J}_\mrm{horizon}}{\eta_H} = \frac{-144 \pi ^2 T^2 \vartheta(z_0)}{18 \alpha  \mu ^2+m^2\ell^2\vartheta(z_0)\rho},
\ee
where 
\be
\rho \equiv 3 \mu ^2+4 \pi  T \left(\sqrt{3 \mu ^2+4 \pi ^2
   T^2}+2 \pi  T\right).
\ee
Eqn. \eqref{Ratiodilat} agrees well with the numerical data in the probe limit.

\begin{figure}
\centering
\includegraphics[width=8.5cm,clip=true]{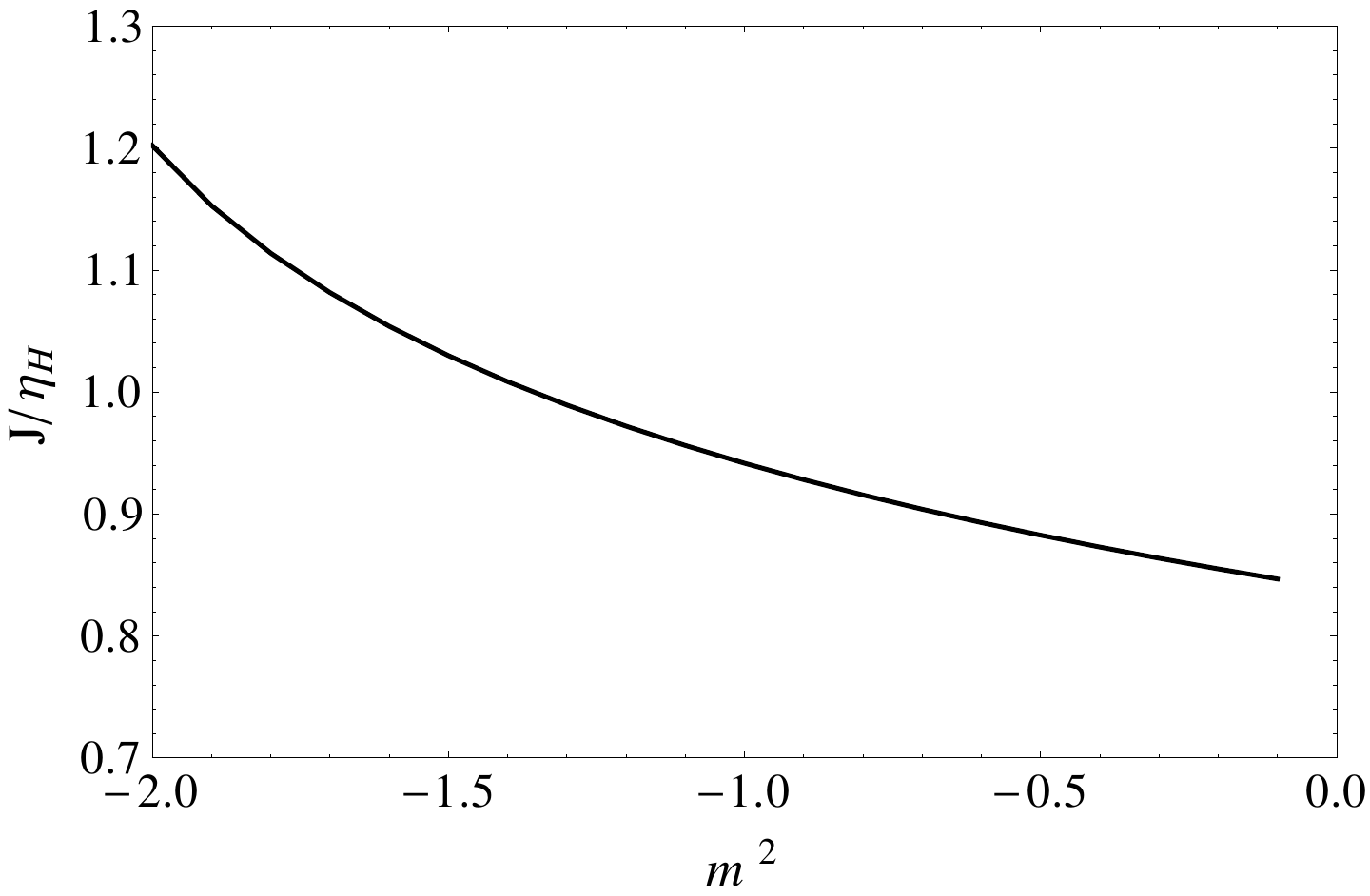}
\caption{(Color online) Angular momentum density to Hall viscosity ratio as a function of $m^2$ for $\mu^2/T^2=0.1$.}
\label{fig:massanalysis_RIIB}
\end{figure}

\section{Discussion}
In this paper, we computed the angular momentum density and the Hall viscosity for specific classes of holographic models 
dual to gapless relativistic quantum systems in $(2+1)$ dimensions. 
Unlike gapped systems at zero temperature, no simple relation is known 
between the angular momentum and the Hall viscosity. We found that 
although the angular momentum density receives contributions both from the
gauge and gravitational Chern-Simons terms, the Hall viscosity requires the gravitational Chern-Simons term. 
This highlights a distinction between the two quantities for gapless systems.

Moreover, when the operator dual to the scalar field $\vartheta$ is marginal, the 
Hall viscosity vanishes even when $\vartheta$ couples to the gravitational ${}^* R R$. 
This is because the holographic expression for the Hall viscosity (\ref{Hallformula})
is proportional to $C' = \partial_z C(\vartheta(z))$ at the horizon $z=z_0$. In order for it to be non-zero, some
energy scale is required. We therefore conjecture that the Hall viscosity $\eta_H$ vanishes
for a conformal field theory, at least in the large $N$ limit.

Nevertheless, we also found that when both quantities are induced by the gravitational Chern-Simons term, 
their ratio shows universal properties as the systems approach their criticalities.
In section III, we studied a boundary system perturbed by a relevant operator of 
dimension $\Delta$. The operator is odd under the parity and time-reversal symmetries, and thus
we are breaking these symmetries explicitly. 
To the leading order in the pertubative expansion, the holographic computation shows that
the ratio of the angular momentum density ${\cal J}$ and the
Hall viscosity $\eta_H$ depends only on $\Delta$ and is given by,
\be
\frac{\cal J}{\eta_H} = \frac{9}{\Delta (3-\Delta)} - \frac{3}{4} + O(3-\Delta).
\ee
The $1/(3-\Delta)$ pole is a reflection of the fact that the Hall viscosity $\eta_H$ vanishes in the
marginal case of $\Delta =3$. 

We also found that the angular momentum density can be decomposed as ${\cal J}= {\cal J}_{{\rm horizon}}+
{\cal J}_{{\rm integral}}$, where ${\cal J}_{{\rm horizon}}$ is a contribution from the horizon, and
${\cal J}_{{\rm integral}}$ is an integral from the horizon to the boundary. On the other hand, $\eta_H$ 
depends only on data at the horizon. From the point of view of the boundary theory,
$\eta_H$ and ${\cal J}_{{\rm horizon}}$ are due to IR physics, while ${\cal J}_{{\rm integral}}$ 
depends only dynamics at all scales. We found that $\eta_H$ and ${\cal J}_{{\rm horizon}}$ are related in
a particularly simple way as
\be
\frac{{\cal J}_{{\rm horizon}}}{\eta_H}= \frac{9}{\Delta (3-\Delta)} ,
\label{exactpole}
\ee
with no $O(3-\Delta)$ corrections. We also found 
similar universal properties when the parity and time reversal symmetries are broken by the dilatonic coupling. 
It would be interesting to find out if the decomposition ${\cal J}= {\cal J}_{{\rm horizon}}+
{\cal J}_{{\rm integral}}$ can be explained from the point of the boundary theory 
and if (\ref{exactpole}) can be derived using its conformal perturbation.

\section*{}
\acknowledgments
We thank N.~Read, O.~Saremi, D.~T.~Son and C.~Wu for useful discussion. 
HO and BS are supported in part by U.S. DOE grant 
DE-FG03-92-ER40701. The work of HO is also
supported in part by 
a Simons Investigator award from the Simons Foundation, the WPI Initiative of MEXT of Japan, and 
JSPS Grant-in-Aid for Scientific Research C-23540285. He also thanks the hospitality of the Aspen Center for Physics and
the National Science Foundation, which supports the Center under Grant No. PHY-1066293, and of the Simons Center for Geometry and Physics. The work of BS is supported in part by a Dominic Orr Graduate Fellowship. BS would like to thank the hospitality of the Kavli Institute for the Physics and Mathematics of the Universe and of the Yukawa Institute for Theoretical Physics. HL is supported in part by funds provided by the U.S. Department of Energy (D.O.E.) under cooperative research agreement DE-FG0205ER41360 and thanks
the hospitality of Isaac Newton Institute for Mathematical Sciences.

\appendix 

\section{Analytic calculation in small $\vt_0$ limit} \label{app:A}

This appendix presents some exact results obtained by solving the scalar field equation in the probe limit for a Schwarzschild black brane. For the metric \eqref{metricSCH} and quadratic potential the scalar field equation
\be
\frac{1}{\sqrt{-g}} \pd_a \lb g^{ab} \sqrt{-g} \pd_b \vartheta(z) \rb - V'(\vartheta) = 0
\ee
can be rewritten as
\be
z^2\left(z^3-z_M^3\right)\vartheta'' + z\left(z^3+2z_M^3\right)\vartheta' + m^2 z_M^3 \vartheta = 0.
\ee
This can be solved analytically and the solution is a sum of two hypergeometric functions. Demanding analyticity at the horizon we obtain
\ba
\frac{\vartheta}{\vartheta_0} &=&  z^{3-\Delta } \, _2F_1\left(1-\frac{\Delta }{3},1-\frac{\Delta
   }{3};2-\frac{2 \Delta }{3};\frac{z^3}{z_M^3}\right)\\
   &-&\frac{ \Gamma \left(2-\frac{2 \Delta }{3}\right) \Gamma^2 \left(\frac{\Delta
   }{3}\right) z^{\Delta } z_M^{3-2 \Delta } \,
   _2F_1\left(\frac{\Delta }{3},\frac{\Delta }{3};\frac{2 \Delta
   }{3};\frac{z^3}{z_M^3}\right)}{\Gamma^2 \left(1-\frac{\Delta }{3}\right)
   \Gamma \left(\frac{2 \Delta }{3}\right)}, \nn
\ea
where the first term is the non-normalizable mode and the second the normalizable response. With this expression we find
\ba
c_\eta &=& -\frac{3^{1-\Delta } \pi ^{\Delta -\frac{3}{2}} m^2 \cot
   \left(\frac{\pi  \Delta }{3}\right) \Gamma \left(\frac{3}{2}-\frac{\Delta
   }{3}\right)}{2^{5-\frac{4 \Delta }{3}}\Gamma \left(1-\frac{\Delta }{3}\right)},\\
c_\mrm{horizon} &=& -\frac{3^{3-\Delta } \pi ^{\Delta -\frac{3}{2}} \cot
   \left(\frac{\pi  \Delta }{3}\right) \Gamma \left(\frac{3}{2}-\frac{\Delta
   }{3}\right)}{2^{5-\frac{4 \Delta }{3}}\Gamma \left(1-\frac{\Delta }{3}\right)},
\ea
and
\ba
& &c_\mrm{integral} = \frac{2^{2 \Delta -7}\pi ^{\Delta -1} (2 \Delta -3)}{3^\Delta \Delta  \Gamma^2\left(1-\frac{\Delta }{3}\right) \Gamma \left(\frac{2 \Delta }{3}\right)} \csc
   \left(\frac{2 \pi  \Delta }{3}\right) \times \nn\\
&\times&  \Bigg\{3 \Gamma^2 \left(\frac{\Delta
   }{3}+1\right) \Gamma \left(\frac{\Delta +4}{3}\right) \Bigg[\Delta  (\Delta +4)\times\nn\\ 
&\times& \, _3  
   \tilde{F}_2\left(\frac{\Delta+3}{3},\frac{\Delta +3}{3},\frac{\Delta +7}{3};\frac{\Delta +10}{3},\frac{2\Delta }{3}+1;1\right) \nn\\
&+& 9 \, _3\tilde{F}_2\left(\frac{\Delta +4}{3},\frac{\Delta
   }{3},\frac{\Delta }{3};\frac{\Delta +7}{3},\frac{2 \Delta
   }{3};1\right)\Bigg]\\
&-& 9 \Delta  \Gamma \left(1-\frac{\Delta
   }{3}\right) \Gamma \left(2-\frac{\Delta }{3}\right) \Gamma
   \left(\frac{7}{3}-\frac{\Delta }{3}\right)\times\nn\\
&\times& \, _3\tilde{F}_2\left(1-\frac{\Delta
   }{3},2-\frac{\Delta }{3},\frac{7}{3}-\frac{\Delta }{3};2-\frac{2 \Delta
   }{3},\frac{10}{3}-\frac{\Delta }{3};1\right)\Bigg\},\nn
\ea
where $\, _3\tilde{F}_2$ is a regularized hypergeometric function defined as
\be
\, _3\tilde{F}_2(a_1,a_2,a_3;b_1,b_2;z) \equiv \frac{_3F_2(a_1,a_2,a_3;b_1,b_2;z)}{\Gamma(b_1)\Gamma(b_2)}.
\ee


\end{document}